\documentclass[preprint]{ptephy}%%%%%% to generate preprint number with ptep logo

\preprintnumber{1509-03386} %%% Insert preprint number here

\usepackage{amsmath}% for dealing with mathematics,
\usepackage{hyperref}% for linking the cross references
\usepackage{graphicx}% for dealing with figures.
\usepackage{subfig}% for getting the subfigures e.g., "Figure 1a and 1b" etc.
\usepackage{dcolumn}% Align table columns on decimal point
\usepackage[mathlines]{lineno}% Enable numbering of text and display math
\usepackage{multirow}
\usepackage{color}

\setcitestyle{numbers}
\begin{document}

\title{Long-lived neutral-kaon flux measurement for the KOTO experiment}

\makeatletter
\let\@fnsymbol\@alph
\makeatother
% Line 104 in ptephy.clo was also changed: \setcounter{footnote}{1} -> \setcounter{footnote}{0}

\author{\name{T.~Masuda}{1,\ast,}\thanks{Present Address : Research Core for Extreme Quantum World, Okayama University, Okayama 700-8530, Japan},
\name{J.~K.~Ahn}{2},
\name{S.~Banno}{3,}\thanks{Present Address : Canon IT Solutions Inc., Tokyo 140-8526, Japan},
\name{M.~Campbell}{4},
\name{J.~Comfort}{5},
\name{Y.~T.~Duh}{6},
\name{T.~Hineno}{1,}\thanks{Present Address : Fujitsu FIP Corporation, Tokyo, Japan},
\name{Y.~B.~Hsiung}{6},
\name{T.~Inagaki}{7},
\name{E.~Iwai}{3,}\thanks{Present Address : High Energy Accelerator Research Organization (KEK), Ibaraki 305-0801, Japan},
\name{N.~Kawasaki}{1},
\name{E.~J.~Kim}{8},
\name{Y.~J.~Kim}{9},
\name{J.~W.~Ko}{9},
\name{T.~K.~Komatsubara}{7},
\name{A.~S.~Kurilin}{10,}\thanks{Deceased},
\name{G.~H.~Lee}{8},
\name{J.~W.~Lee}{3,}\thanks{Present Address : Department of Physics, Korea University, Seoul 136-713, Republic of Korea},
\name{S.~K.~Lee}{8},
\name{G.~Y.~Lim}{7},
\name{J.~Ma}{11,}\thanks{Present Address : Mayo Clinic, Rochester, MN 55905, USA},
\name{D.~MacFarland}{5},
\name{Y.~Maeda}{1,}\thanks{Present Address : Kobayashi-Maskawa Institute, Nagoya University, Nagoya 464-8602, Japan},
\name{T.~Matsumura}{12},
\name{R.~Murayama}{3},
\name{D.~Naito}{1},
\name{Y.~Nakaya}{3,}\footnotemark[5],
\name{H.~Nanjo}{1},
\name{T.~Nomura}{7},
\name{Y.~Odani}{13,}\thanks{Present Address : JASTEC Co., Ltd., Tokyo 108-0074, Japan},
\name{H.~Okuno}{7},
\name{Y.~D.~Ri}{3,}\thanks{Present Address : Hitachi Industry and Control Solutions, Ltd., Ibaraki 319-1221, Japan},
\name{N.~Sasao}{14},
\name{K.~Sato}{3,}\thanks{Present Address : Institute of Cosmic Ray Research, University of Tokyo, Chiba 277-8582, Japan},
\name{T.~Sato}{7},
\name{S.~Seki}{1},
\name{T.~Shimogawa}{13,}\footnotemark[4],
\name{T.~Shinkawa}{12},
\name{K.~Shiomi}{3,}\footnotemark[4],
\name{J.~S.~Son}{8},
\name{Y.~Sugiyama}{3},
\name{S.~Suzuki}{13},
\name{Y.~Tajima}{15},
\name{G.~Takahashi}{1},
\name{Y.~Takashima}{3,}\thanks{Present Address : Gigaphoton Inc., Tochigi 323-8558, Japan},
\name{M.~Tecchio}{4},
\name{M.~Togawa}{3},
\name{T.~Toyoda}{3,}\thanks{Present Address : Hitachi, Ltd., Tokyo 100-8280, Japan},
\name{Y.~C.~Tung}{6,}\thanks{Present Address : Enrico Fermi Institute, University of Chicago, Chicago, IL 60637, USA},
\name{Y.~W.~Wah}{11},
\name{H.~Watanabe}{7},
\name{J.~K.~Woo}{9},
\name{J.~Xu}{4},
\name{T.~Yamanaka}{3},
\name{Y.~Yanagida}{3,}\thanks{Present Address : Shimadzu Corporation, Kyoto 604-8511, Japan},
\name{H.~Y.~Yoshida}{15}, and
\name{H.~Yoshimoto}{3,}\thanks{Present Address : Resona Bank, Limited., Osaka 540-8610,Japan}
}
%%%%%%%%%%% The \name command should be used as \name{Insert author name here}{Insert affiliation number here}
%%%%% Please use \thanks for contributed author details

%%%%%%%%%%% The \affil command should be used as \affil{Insert affiliation number here}{Insert author address here}
\address{\affil{1}{Department of Physics, Kyoto University, Kyoto 606-8502, Japan}
\affil{2}{Department of Physics, Korea University, Seoul 136-713, Republic of Korea}
\affil{3}{Department of Physics, Osaka University, Osaka 560-0043, Japan}
\affil{4}{Department of Physics, University of Michigan, Ann Arbor, MI 48109, USA}
\affil{5}{Department of Physics, Arizona State University, Tempe, AZ 85287, USA}
\affil{6}{Department of Physics, National Taiwan University, Taipei, Taiwan 10617, Republic of China}
\affil{7}{High Energy Accelerator Research Organization (KEK), Ibaraki 305-0801, Japan}
\affil{8}{Division of Science Education, Chonbuk National University, Jeonju 561-756, Republic of Korea}
\affil{9}{Department of Physics, Jeju National University, Jeju 690-756, Republic of Korea}
\affil{10}{Laboratory of Nuclear Problems, Joint Institute for Nuclear Researches, Dubna, Moscow reg. 141980, Russia}
\affil{11}{Enrico Fermi Institute, University of Chicago, Chicago, IL 60637, USA}
\affil{12}{Department of Applied Physics, National Defense Academy, Kanagawa 239-8686, Japan}
\affil{13}{Department of Physics, Saga University, Saga 840-8502, Japan}
\affil{14}{Research Core for Extreme Quantum World, Okayama University, Okayama 700-8530, Japan}
\affil{15}{Department of Physics, Yamagata University, Yamagata 990-8560, Japan}
\email{masuda@okayama-u.ac.jp}
}

\begin{abstract}%
The KOTO ($K^0$ at Tokai) experiment aims to observe the CP-violating rare decay $K_L \rightarrow \pi^0 \nu \bar{\nu}$ by using a long-lived neutral-kaon beam produced by the 30~GeV proton beam at the Japan Proton Accelerator Research Complex. The $K_L$ flux is an essential parameter for the measurement of the branching fraction. Three $K_L$ neutral decay modes, $K_L \rightarrow 3\pi^0$, $K_L \rightarrow 2\pi^0$, and $K_L \rightarrow 2\gamma$ were used to measure the $K_L$ flux in the beam line in the 2013 KOTO engineering run. A Monte Carlo simulation was used to estimate the detector acceptance for these decays. Agreement was found between the simulation model and the experimental data, and the remaining systematic uncertainty was estimated at the 1.4\% level. The $K_L$ flux was measured as $(4.183 \pm 0.017_{\mathrm{stat.}} \pm 0.059_{\mathrm{sys.}}) \times 10^7$ $K_L$ per $2\times 10^{14}$ protons on a 66-mm-long Au target.
\end{abstract}

\subjectindex{C30, G12}

\maketitle

%% Start line numbering here
%\linenumbers

\section{\label{sec:intro}Introduction}
The KOTO ($K^0$ at Tokai) experiment is aimed at the first observation of the rare CP-violating decay of the long-lived neutral-kaon $K_L \rightarrow \pi^0 \nu \bar{\nu}$. An important characteristic of this decay is its theoretical cleanness. The branching fraction (BF) is predicted to be $2.4 \times 10^{-11}$ with a theoretical uncertainty of only 2.5\%~\cite{PhysRevD.83.034030}, in contrast to that of most other meson decays, especially in the $B$ system, where the uncertainty can be as high as 10\%. The study of $K_L \rightarrow \pi^0 \nu \bar{\nu}$ can also shed light on the origin of CP violation as it provides a direct precise measurement of the parameter $\eta$~\cite{wolfenstein}, the height of the unitarity triangle, in the standard model (SM) of particle physics. In addition, $K_L \rightarrow \pi^0 \nu \bar{\nu}$ involves an $s \rightarrow d$ transition that is highly suppressed in the SM and thus is very effective in searching for new physics that modifies the SM flavor structure. The current experimental upper limit on the BF is $2.6 \times 10^{-8}$ at the 90\% confidence level, obtained by the KEK-E391a experiment~\cite{Ahn:2010fk} at the KEK 12~GeV proton synchrotron (KEK-PS).

The KOTO experiment~\cite{Yamanaka01012012} uses the experimental technique established by E391a. In order to improve the sensitivity by three orders of magnitude, a new dedicated beam line was constructed at the Japan Proton Accelerator Research Complex (J-PARC)~\cite{Nagamiya01012012}. The design value of the proton intensity delivered by the Main Ring (MR) at J-PARC is 100 times higher than that of the KEK-PS. To improve the background suppression, the main electromagnetic calorimeter and other detectors surrounding the decay volume were upgraded.

The $K_L$ flux generated by the proton beam is an essential parameter for the measurement of the BF of $K_L \rightarrow \pi^0 \nu \bar{\nu}$. The KOTO Collaboration measured the $K_L$ flux and the momentum spectrum by reconstructing the charged decay modes $K_L \rightarrow \pi^+ \pi^- \pi^0$ and $K_L \rightarrow \pi^+ \pi^-$ in beam surveys with dedicated tracking detectors and calorimeters~\cite{Shiomi2012264,dt_sato}. During data acquisition, the $K_L$ flux is measured using the neutral decay modes $K_L \rightarrow 3\pi^0$, $K_L \rightarrow 2\pi^0$, and $K_L \rightarrow 2\gamma$, since the KOTO detector cannot measure the momentum of charged pions. In this paper, we report on KOTO's first measurement of the $K_L$ flux by using the three neutral decay modes with the main calorimeter, which was performed during an engineering run in January 2013.

The paper is organized as follows. Section~\ref{sec:apparatus} describes the experimental setup relevant to the measurement. Section~\ref{chap:fluxmeasurement} illustrates the principle and the technique used in the measurement. Section~\ref{sec:mc} reports on the Monte Carlo (MC) simulation used to measure the acceptance of each of the three modes. Section~\ref{chap:analysis} describes the analysis of the experimental data for each of the three modes. Section~\ref{sec:kl_flux} reports the results and their uncertainty estimations. Section~\ref{sec:discussion} compares this to previous results. Section~\ref{sec:conclusion} describes the conclusions.

\section{Experimental setup}\label{sec:apparatus}
This section focuses on the experimental setup used for the measurement, including descriptions of the beam line and of the subsystems in the KOTO detector relevant for this analysis.

\subsection{\label{sec:beam}Beam Line}
\begin{figure}
  \centering
  \includegraphics[width=14cm, viewport=0 0 1000 550, clip]{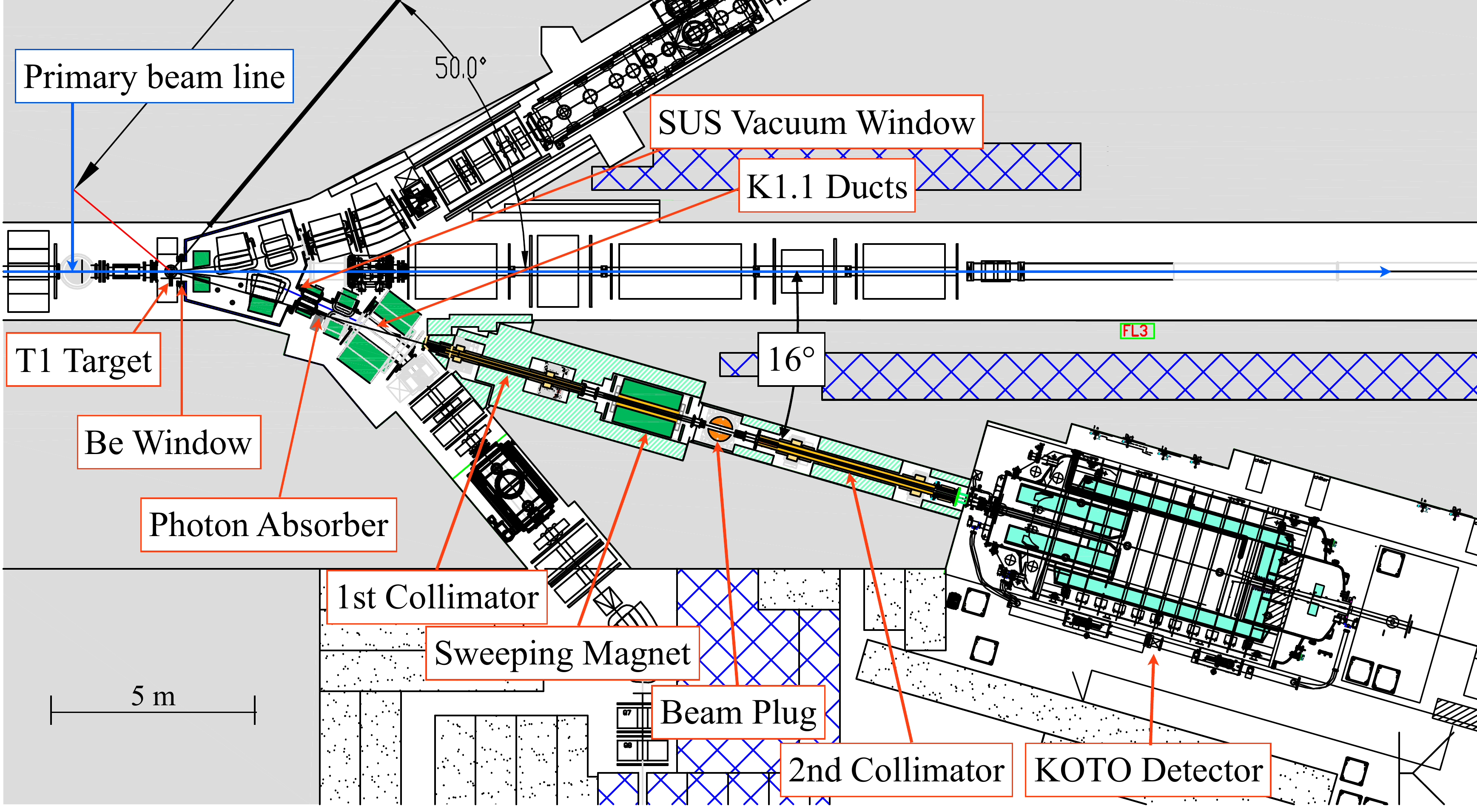}
  \caption{Schematic top view of the KL beam line used by the KOTO experiment. The primary proton beam comes in from the left. The KL beam line starts at the T1 target and the beam axis goes off at an angle of 16$^\circ$ with respect to the primary beam line. The front of the KOTO detector is located at the end of the beam line, about 21 m from the T1 target.}
  \label{fig:secondarybeamline}
 \end{figure}

The KOTO detector is located at the Hadron Experimental Facility of J-PARC, at the end of a new 21-m-long neutral-kaon beam line (KL). A schematic top view of the KL beam line and the KOTO detector is shown in Fig.~\ref{fig:secondarybeamline}. The protons are accelerated to 30~GeV by the MR, extracted by using a slow extraction technique~\cite{Koseki}, and transported through the primary beam line to the facility~\cite{Agari01012012}. The proton beam intensity is monitored by a secondary emission chamber (SEC), located after the extraction point. The primary proton beam, with a cross section of approximately 1~mm in radius, is injected into the production target (T1). The target consists of a 66-mm-long gold target of $6\times 6~\mathrm{mm^2}$ cross section. It is equally divided into six parts along its length with five 0.2-mm-thick slits.

\begin{figure}
  \centering
  \includegraphics[width=15cm, viewport=0 0 1050 350, clip]{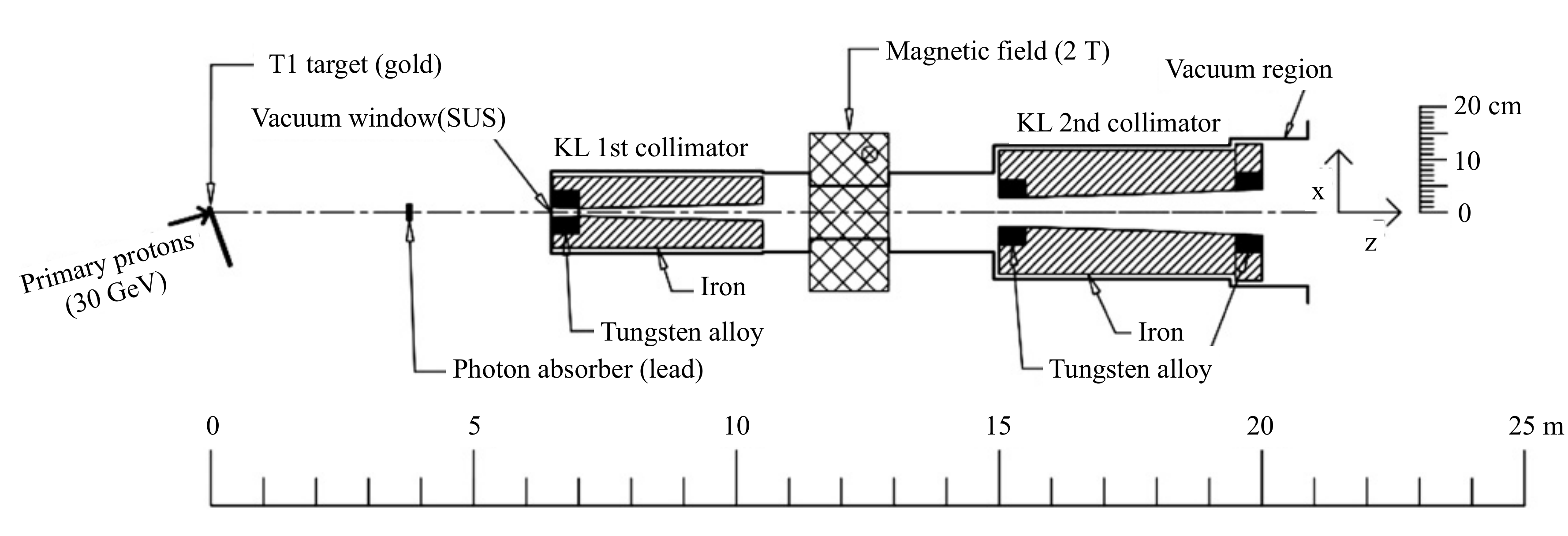}
  \caption{KOTO collimation scheme in the beam line coordinates \cite{Shimogawa2010585}.}
  \label{fig:collimators}
 \end{figure}

 The KL beam line is off-axis by an angle of 16$^\circ$ with respect to the primary proton beam line. The full kinematic reconstruction of the neutral pion in the $K_L \rightarrow \pi^0 \nu \bar{\nu}$ decay requires a small diameter for the $K_L$ beam and was achieved with the collimation scheme shown in Fig.~\ref{fig:collimators}. The secondary particles produced at the target pass through the pair of collimators shown in Fig.~\ref{fig:collimators} to shape the beam. At the exit of the second collimator, the neutral beam has a square cross-section of $8.5 \times 8.5~\mathrm{cm^2}$ corresponding to a solid angle of 7.8~$\mu$sr. Before entering the collimation region, the beam passes through a 7-cm-thick (12.5~$X_0$) lead absorber which removes most of the photons. A 2-T dipole magnet, located between the two collimators sweeps out charged particles. Short-lived particles decay in the long collimator region leaving only $K_L$ mesons, neutrons, and photons at the entry of the detector region. Table~\ref{tab:materialsinbeam} lists the composition and location of all the components along the KL beam line. The materials in the beam decrease the $K_L$ flux by 60\%, as estimated using a Geant4~\cite{Agostinelli2003250, 1610988} based MC simulation. The collimation scheme of the KL beam line is described in Ref.~\cite{Shimogawa2010585}.
 
\begin{table}
\caption{Composition and position of materials along the axis of the KL beam line: some components are part of the K1.1 charged-kaon beam line, which uses the same T1 target and is at an angle of 7$^\circ$ with respect to the KL beam line. The thickness is listed only for materials actually crossed by the beam. The regions between the Be vacuum window and the stainless steel vacuum window, and between the K1.1 front duct and the beam exit window are in a 2-Pa vacuum. The starting position is in beam line coordinates, whose origin is at the center of the T1 target. The KOTO detector starts at the front face of the Front Barrel; detector coordinates are measured with respect to this position.}
\begin{center}
\begin{tabular}{lcccc} \hline \hline
\multirow{2}{*}{Name} & \multirow{2}{*}{Material} & Thickness & Starting position \\
                                        &                                             & [mm]           & [mm] & \\
\hline
T1 target & Au & 66 & -   \\
Vacuum window & Be & 8 & 247 \\ 
Vacuum window & Stainless steel & 0.2 & 3,097  \\
 Photon absorber & Pb & 70 & 3,730  \\
 K1.1 front duct & Stainless steel  & 0.2 & 4,182   \\
 K1.1 tail duct & Stainless steel  & 0.2 & 5,510  \\
Collimator vacuum window & Stainless steel & 0.1 & 6,400  \\
 1$^{\rm st}$ collimator & Fe and W alloy & - & 6,500  \\
  2$^{\rm nd}$ collimator & Fe and W alloy & - & 15,000  \\
Beam exit vacuum window & Polyimide & 0.125 & 20,000  \\
Front Barrel &  -   & - & 21,507\\
CsI calorimeter &   & - & 27,655  \\
\hline \hline
\end{tabular}
\end{center}
\label{tab:materialsinbeam}
\end{table}%

\subsection{\label{sec:detectors}KOTO detector}
\begin{figure}
 \centering
 \includegraphics[width=16.5cm, bb=20 0 770 210, clip]{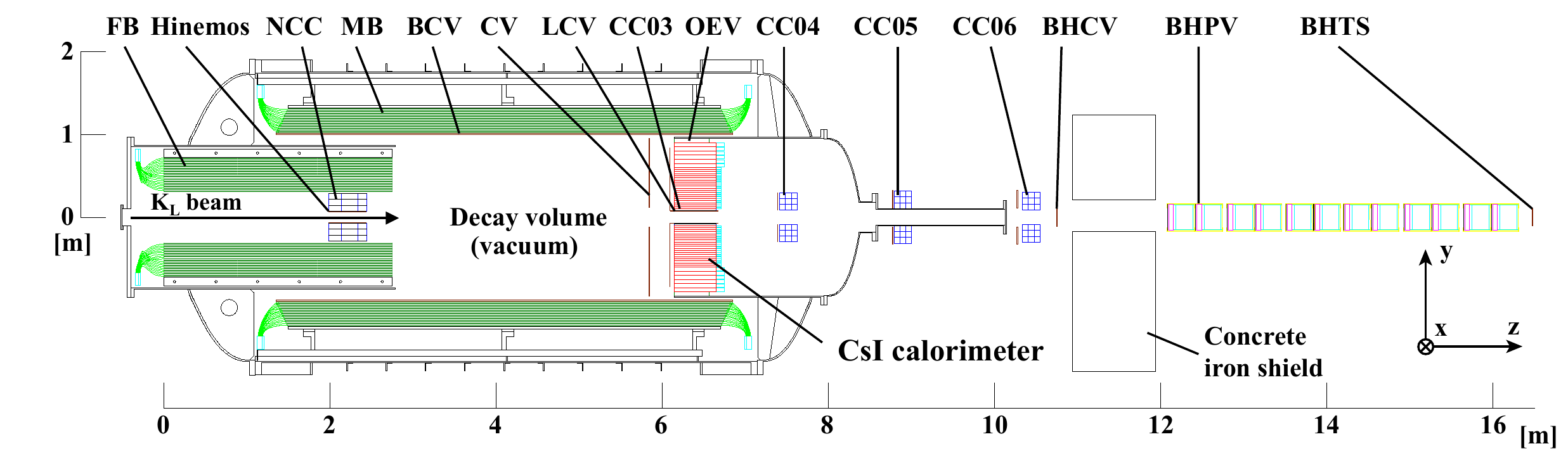}
 \caption{Cross-sectional side view of KOTO in detector coordinates, defined with respect to the front face of the Front Barrel (FB) detector. $K_L$ particles come in from the left. The $z$-axis is along the beam direction, the $y$-axis points upward, and the $x$-axis goes into the page. The CsI calorimeter, located downstream of the decay volume, is surrounded by two photon detectors (FB and MB), charged-particle veto detectors (CV, LCV, BCV, and Hinemos), photon veto collar counters made of CsI crystals (NCC, CC03, CC04, CC05, and CC06) and forward charged-particle and photon detectors (BHCV and BHPV). The gap between the MB and the CsI calorimeter is filled with a photon detector (OEV). BHTS is a trigger scintillator which is used for detector performance checks.}
 \label{fig:cutoffview}
 \end{figure}

  Figure~\ref{fig:cutoffview} shows a cross-sectional side view and the coordinate system of the KOTO detector. The signature of a $K_L \rightarrow \pi^0 \nu \bar{\nu}$ decay is a pair of photons coming from the $\pi^0$ decay, without any other detectable particles. The energy and position of the two photons are measured with a cesium iodide electromagnetic calorimeter (CsI)~\cite{nim_iwai}. Multiple charged-particle and photon detectors surround the decay volume to form a hermetic veto against any extra particles except neutrinos. The decay vertex of the $K_L$ is reconstructed under the assumption that the two photons come from a $\pi^0$ on the beam axis and that the vertices of the $K_L$ and $\pi^0$ coincide. Finally the $\pi^0$ is required to have a large transverse momentum to balance the momentum carried by the two neutrinos.
  
  For this $K_L$ flux measurement, in addition to the CsI calorimeter, the Main Barrel (MB)~\cite{Tajima2008261} and the Charged Veto (CV)~\cite{maeda_pic}, which are described below, are used as veto detectors. 
    
The CsI calorimeter consists of 2716 undoped CsI crystals stacked in a cylindrical shape of 2-m diameter and 500-mm depth (27~$X_0$) along the beam direction. We use crystals of two sizes: 2240 small crystals of $25 \times 25~\mathrm{mm^2}$ cross section for the inner region and 476 large crystals of $50 \times 50~\mathrm{mm^2}$ for the outer region. The calorimeter has a $148 \times 148~\mathrm{mm^2}$ square beam hole at its center. The CsI crystals have a typical light yield of 9~photoelectrons~(p.e.)/MeV, and an energy resolution $\sigma_E/E$ of $1.9\%/\sqrt{E~\mathrm{[GeV]}}$ with a constant term of 0.6\%. They are read out by photomultiplier tubes (PMTs)~\cite{Masuda201411}.

\begin{figure}
 \centering
 \includegraphics[width=9cm, bb=0 0 720 640, clip]{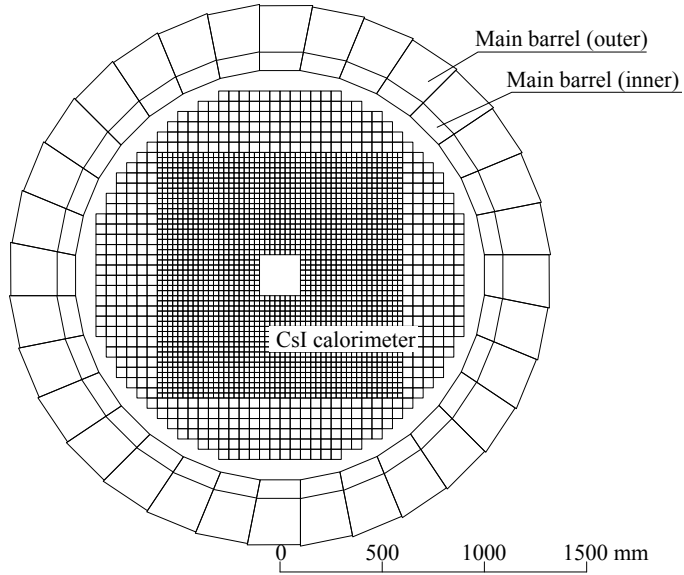}
 \caption{Cross-sectional end view of the CsI calorimeter and the MB photon veto detector.}
 \label{fig:csi_mb_rear}
 \end{figure}

The MB is a cylindrical photon veto detector 5.5~m long with 2-m inner diameter surrounding the decay volume. It consists of 32 modules, each made of 45 alternate layers of plastic scintillator and lead sheet. The inner 15 layers have 1-mm-thick lead sheets while the outer 30 layers have 2-mm-thick lead sheets. The thickness of the plastic scintillator is 5~mm for both the inner and outer layers. The total thickness of the modules is 13.5~$X_0$. The scintillation light originating in the plastic scintillator is absorbed by wavelength-shifting fibers (WLSFs) and read out by PMTs that are connected at both ends of each inner and outer module. Figure~\ref{fig:csi_mb_rear} shows a cross-sectional end view of the CsI calorimeter and the MB photon veto detector.

\begin{figure}
 \centering
 \includegraphics[width=16cm, bb=0 0 1000 500, clip]{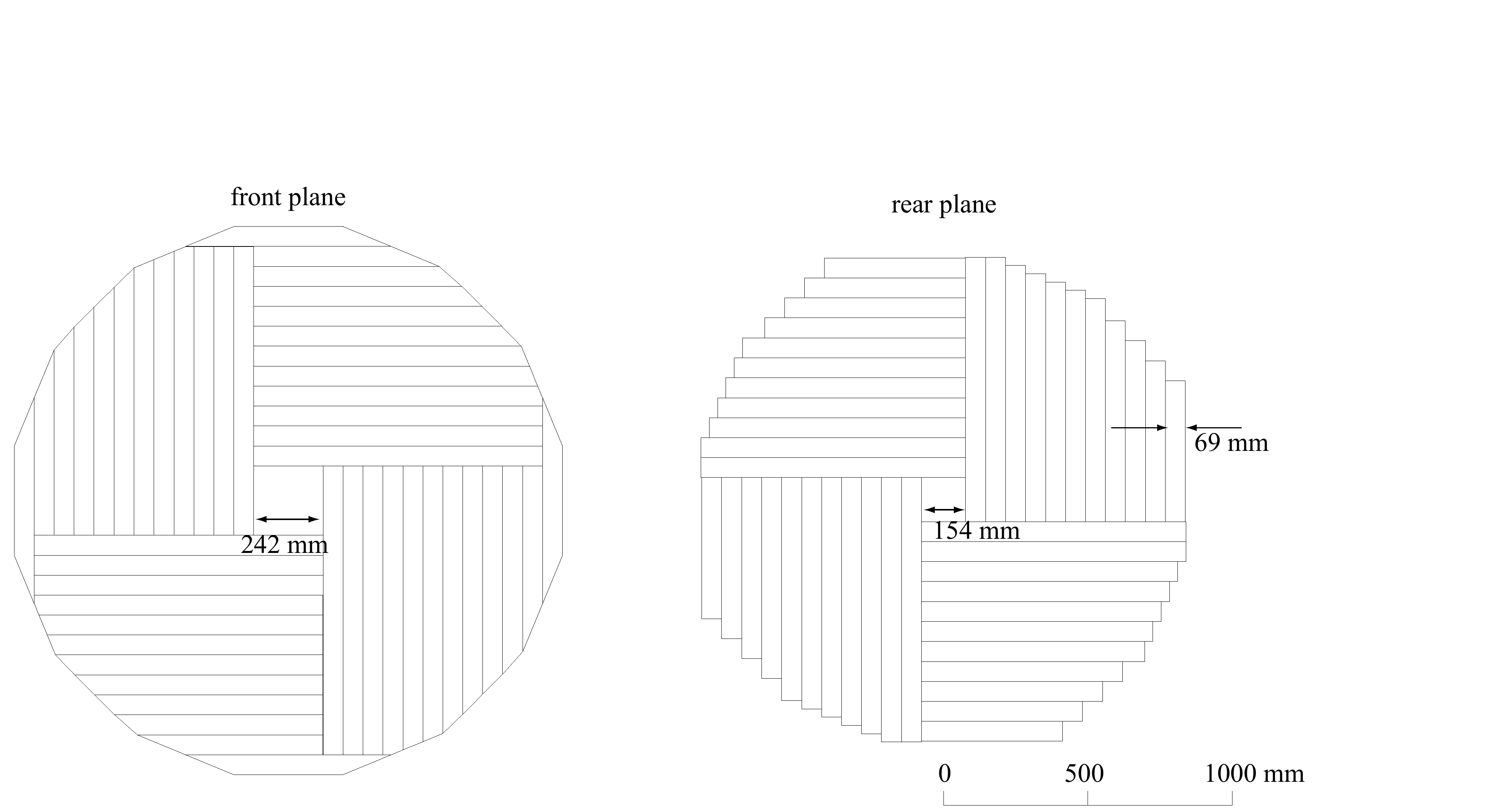}
 \caption{End view of the CV detector showing the length and orientation of the 69 mm-wide plastic scintillator strips in each of the two CV layers.}
 \label{fig:cv}
 \end{figure}

The CV is the main charged-particle veto detector. It consists of two layers, placed 5~cm and 30~cm upstream of the calorimeter. Each layer is made of 3-mm-thick and 69-mm-wide plastic scintillator strips of lengths between 490 and 1002~mm, assembled as shown in Fig.~\ref{fig:cv}. The scintillation light is picked up by WLSFs and read out by multi pixel photon counters at both ends of each strip. The typical light yield of the strips is 186~p.e./MeV and the typical time resolution is 1.2~ns when averaging over the two ends~\cite{naito}.

These three detectors are all contained inside a pressure vessel defining the decay volume, which is evacuated to $10^{-3}$~Pa to suppress interactions of the beam particles with the residual gas. Due to a large amount of outgassing, the detectors inside the pressure vessel are separated from the high-vacuum region by a thin membrane and evacuated at a level of 1~Pa.
 
 Analog signals from all detectors are digitized by ADC modules. A filter inside the ADC modules shapes the raw front-end signals into Gaussian pulses approximately 50~ns wide before digitizing them at a 125~MHz sampling rate and with 14~bit resolution~\cite{Bogdan:2007kq}. The digitized waveforms are transmitted to the data acquisition (DAQ) system via optical links and processed by a three-tiered trigger system~\cite{Tecchio20121940}. The waveform consists of 64 consecutive 8-ns samples around the signal region, corresponding to a 512-ns time window.

\section{$\mathrm{K_L}$ flux measurement}
\label{chap:fluxmeasurement}
 We performed an engineering run in January 2013 to check the performance of the KOTO detector, including the data readout and peripheral slow control systems. In addition to the commissioning of the detector, the $K_L$ beam flux with the detector in situ was measured. This section describes the principles of the $K_L$ flux measurement, the run conditions under which the data were taken, and the triggers used during the run.

\subsection{Measurement principle}\label{sec:principle}
 We determined the $K_L$ flux at the beam exit vacuum window (see Table.~\ref{tab:materialsinbeam}). The number of $K_L$s were counted by using the neutral decay modes, $K_L \rightarrow 3\pi^0$, $K_L \rightarrow 2\pi^0$, and $K_L \rightarrow 2\gamma$, which have large statistics. Their branching fractions are listed in Table~\ref{tab:klbr}. These decay modes have a common final-state configuration with all the decay products being photons. They provide three independent results that can be compared and cross-checked against each other. 
 
\begin{table}
 \caption{Branching fractions (BFs) of the three $K_L$ neutral decay modes~\cite{pdg} used for the $K_L$ flux measurement.} \label{tab:klbr}
 \centering
 \begin{tabular}{lc} \\ \hline
    Mode & BF \\ \hline
    $K_L \rightarrow 3\pi^0$ & $0.1952 \pm 0.0012$  \\
    $K_L \rightarrow 2\pi^0$ & $(8.64 \pm 0.06) \times 10^{-4}$  \\
    $K_L \rightarrow 2\gamma$ & $(5.47 \pm 0.04) \times 10^{-4}$ \\ \hline
 \end{tabular}
\end{table}
 
 The $K_L \rightarrow 2\gamma$ decays were reconstructed from events with only two photons assuming the nominal $K_L$ mass. The $K_L \rightarrow 2\pi^0$ and $K_L \rightarrow 3\pi^0$ decays were reconstructed from two orthogonal sets of events with exactly four and six photons detected in the CsI calorimeter, respectively. To reconstruct the $\pi^0$s, all possible combinations of two photon pairs were considered. The decay vertex position was calculated from the energy and position of the two photons in each pair, assuming the nominal $\pi^0$ mass. The photon pairing was determined by minimizing the variance of the common vertex position, as described in Sect.~\ref{sec:vertexreconstruction}.

The MB inner layers were used as a photon veto to suppress the $K_L \rightarrow 3\pi^0$ background in the $K_L \rightarrow 2\pi^0$ and $K_L \rightarrow 2\gamma$ samples, when some of the photons fell outside the CsI fiducial area. The CV was used as a charged particle veto to remove backgrounds due to the $K_L \rightarrow \pi^{\pm} e^{\mp} \nu$ ($K_{e3}$) and $K_L \rightarrow \pi^+ \pi^- \pi^0$ decays by identifying  charged-particles before they hit the CsI calorimeter.

\subsection{Run condition}\label{sec:runcondition}
The data used for this analysis were accumulated during a 6-hour run taken on January 16, 2013. The nominal beam power was 15~kW, corresponding to an intensity of $1.8\times 10^{13}$ protons on target (POT) per spill.  A spill had a duration of 2~s and a repetition rate of 6~s. The standard deviation of the beam intensity spill by spill was 0.7\%. The total POT during the run, as measured by SEC, was $(3.355 \pm 0.013) \times 10^{16}$. The uncertainty was attributable mainly to the conversion factor from the SEC counts to POT.

\subsection{Trigger}\label{sec:trigger}
The $K_L$ neutral decays used in the $K_L$ flux measurement were triggered by requiring that the energy depositions in both the left and right halves of the CsI calorimeter be higher than a threshold. The common threshold for both halves of the calorimeter was 307.5~MeV, which resulted in a trigger rate of about 34~kHz. This trigger scheme is effective at selecting events with low missing energy.

\section{Monte Carlo simulation}\label{sec:mc}

 \begin{figure}
  \centering
  \includegraphics[width=11cm, bb=0 0 550 350, clip ]{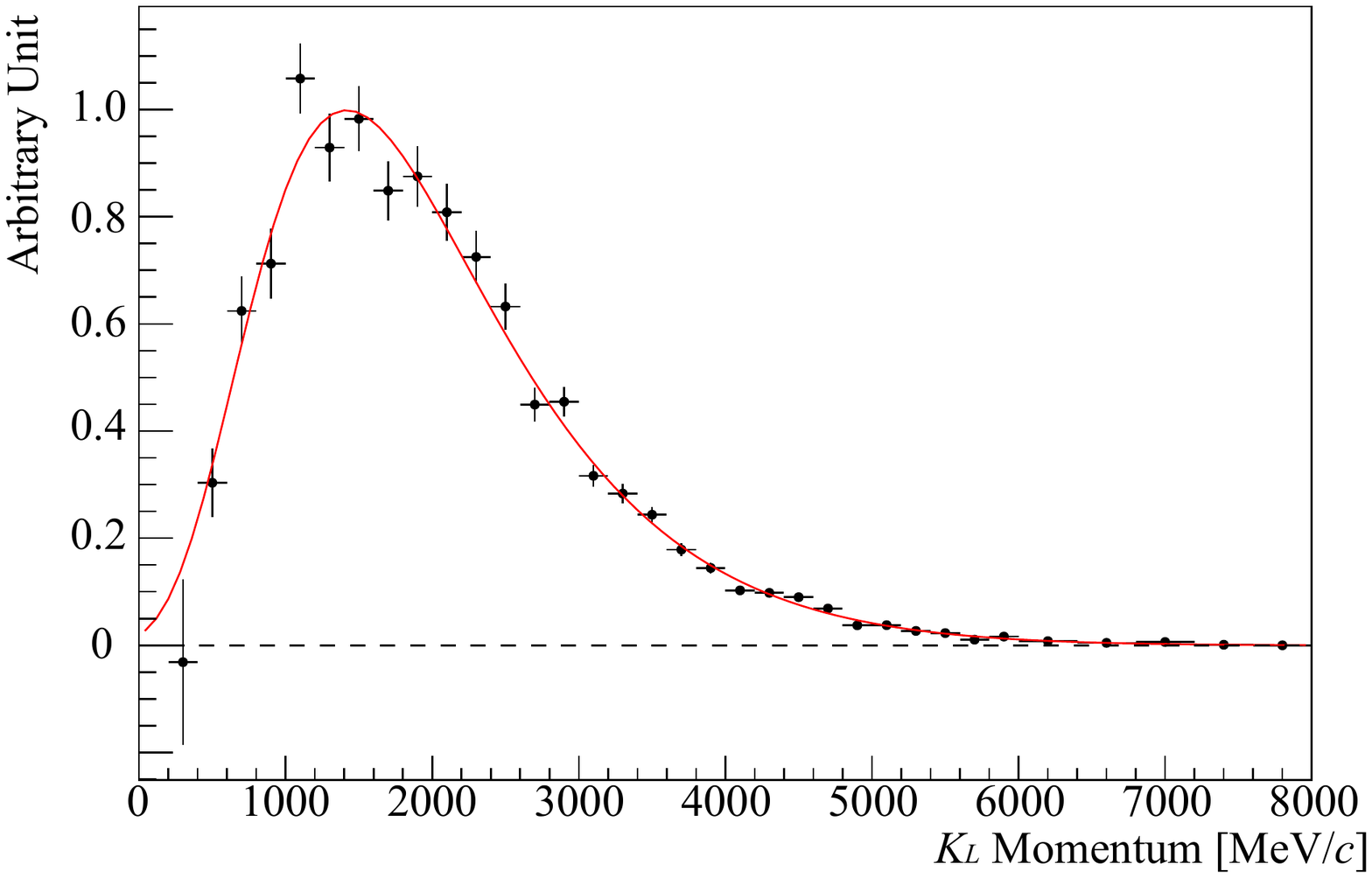}
  \caption{$K_L$ momentum distribution at the beam exit. Points with error bars are experimental data taken with a magnetic spectrometer~\cite{dt_sato}. The spectrum below 2000~MeV/$c$ was measured with $K_L \rightarrow \pi^+ \pi^-$; that above 2000~MeV/$c$ was measured with $K_L \rightarrow \pi^+ \pi^-$ and $K_L \rightarrow \pi^+ \pi^- \pi^0$. The red curve is a Gaussian fit of the data with a momentum-dependent resolution and it is used to generate the $K_L$ spectrum in the MC simulation. The fit returns a peak momentum of 1.42~GeV/$c$.}
  \label{fig:satospectrum}
 \end{figure}

 An MC simulation was used to estimate the acceptances of the detector for the $K_L \rightarrow 3\pi^0$, $K_L \rightarrow 2\pi^0$, and $K_L \rightarrow 2\gamma$ decays. It was also used to evaluate the contamination of background events from other $K_L$ decays. The Geant4 MC package (Geant4.9.5.p02) was used to simulate the geometry of the KOTO detector and the interactions between particles and materials inside the detector.  $K_L$ particles were generated at the beam exit  with the momentum spectrum shown in Fig.~\ref{fig:satospectrum}. The spectrum was measured independently in a special run with a magnetic spectrometer~\cite{dt_sato}.

The detector response was corrected by using data taken in dedicated beam tests, bench tests with a $^{137}$Cs radioactive source, and cosmic-ray runs with the detector in situ.
For the CsI calorimeter, we added to the MC simulation the measured light yield of each crystal and its nonuniformity along the depth of the crystal, the accuracy of the energy calibration, the pedestal fluctuations, the light propagation velocity along the depth of the crystal, and the time smearing due to the timing resolution. For the MB photon veto, the length of the detector resulted in a substantial dependence of the light yield and timing delay on the position of the hit along the module. This effect was simulated by using a light propagation velocity and light attenuation measured in data (see Sect.~\ref{sec:veto}). For the CV, the light yield and timing resolution measured for each strip were used in the simulation.

\subsection{Accidental overlay}\label{sec:overlay}
 The overlap of accidental activity generated by simultaneous $K_L$ decays or interactions of other beam particles was not simulated in the MC. Both of these accidental sources can result in additional energy deposition and possible distortion of the timing information recorded by the front-end electronics. 
To simulate such effects, we overlaid data collected with the accidental trigger described below to MC events passing the same selection criteria as the online trigger logic. The accidental data consisted of events triggered with the same trigger configuration described in Sect.~\ref{sec:trigger}, but recorded in a time window starting 512~ns or 768~ns before the actual trigger.
The energy deposition measured in a randomly chosen accidental event was added to the energy deposition of a simulated event at the level of the ADC readout. If the energy of the overlapping event was larger that the energy of the simulated event, it determined the event timing at the detector.

\section{Analysis}
\label{chap:analysis}
This section describes the calibration, the event reconstruction, and the event selection used for the $K_L$ flux measurement. The photon, $\pi^0$, and $K_L$ reconstruction methods are presented. A more detailed description can be found in Ref.~\cite{dt_masuda}.

\subsection{CsI calibration}\label{sec:calibration}
\begin{figure}
 \centering
 \includegraphics[width=9cm, bb=0 0 500 380, clip]{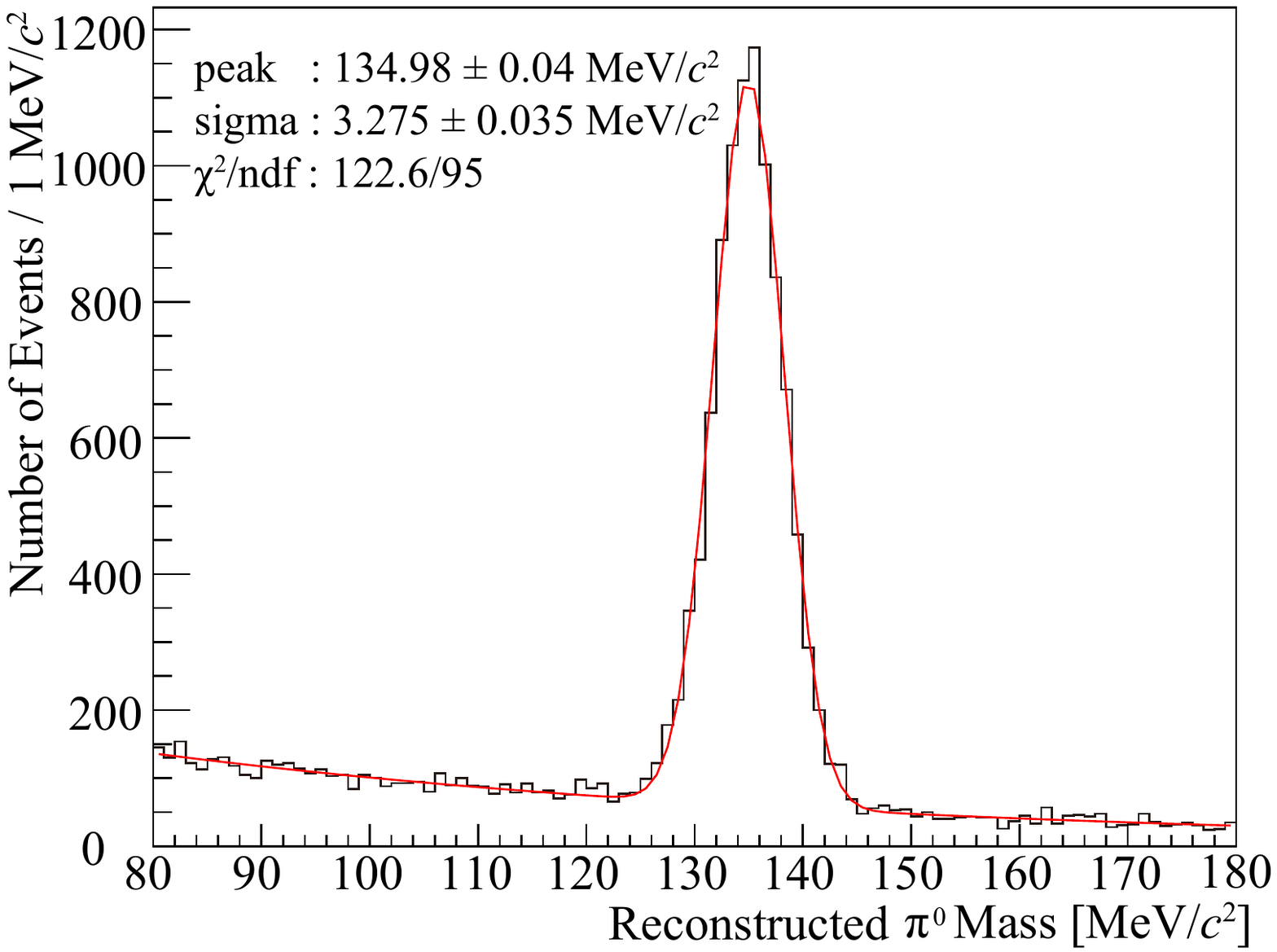}
 \caption{Reconstructed invariant mass of $\pi^0$ events collected in the aluminum plate runs. The distribution is fitted using a Gaussian peak plus an exponential slope. The final fit is shown by the red curve.}
 \label{fig:altargetpeak}
\end{figure}

The energy calibration of the CsI calorimeter was performed in three steps. An initial calibration factor to convert ADC counts into energy was calculated by using vertical cosmic rays penetrating the calorimeter and assuming an energy deposit per unit path length in cesium iodide of 5.6~MeV/cm~\cite{pdg}.
Next, from a large statistics $K_L \rightarrow 3\pi^0$ sample, we tuned the calibration factor of each crystal by minimizing the overall width of the reconstructed $K_L$ invariant mass distribution~\cite{jwlee}. Finally, an absolute correction to the calibration factor was obtained by using events with $\pi^0$ of known decay positions. These data were collected in special runs with a 5~mm-thick aluminum plate placed at 3353~mm upstream of the CsI front face. Neutrons in the beam core interacted with the aluminium nuclei and generated $\pi^{0}$ which decayed immediately. The deposit energies of the two photons from the $\pi^0 \rightarrow 2\gamma$ decays were measured by the CsI calorimeter. We scaled the absolute calibration factor of whole crystals to make the invariant mass of the two photons the same as the nominal $\pi^0$ mass. The invariant mass distribution of these events is shown in Fig.~\ref{fig:altargetpeak}.

\subsection{Clustering}\label{sec:clustering}
 \begin{figure}
  \centering
  \includegraphics[width=12cm, bb=0 0 1000 580, clip ]{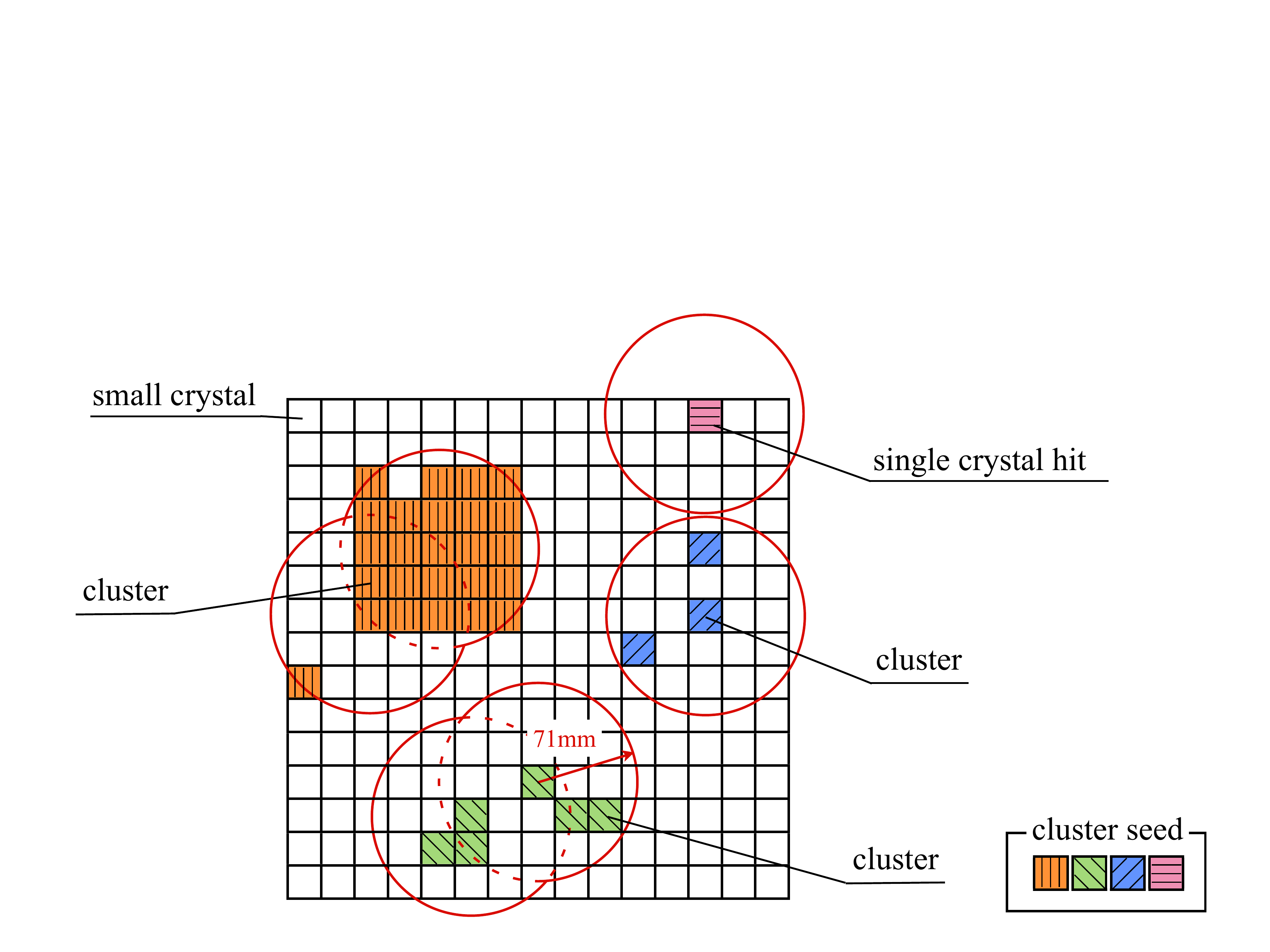}
  \caption{Graphical view of the KOTO clustering algorithm: the colored boxes represent seed crystals. Cluster seeds with the same color share a cluster. The red circles show the cluster-growing radius of 71~mm.}
  \label{fig:clustering}
 \end{figure}
 
A photon hitting the CsI calorimeter induces an electromagnetic shower with a Moli\`{e}re radius of 3.53~cm for cesium iodide~\cite{anp}, thus depositing energy over multiple crystals. First, we selected CsI crystals with energy deposits above 3~MeV and within 150~ns of each other. Such crystals were called ``cluster seeds''. Next we grouped seeds into a single cluster if found within 71~mm of another seed, as shown in Fig.~\ref{fig:clustering}. Once no other seed could be added, we calculated the energy, coordinates, and timing of the cluster as follows:
\begin{eqnarray}
 E_{\mathrm{cluster}} &=& \sum_{i=1}^n e_i, \label{eq:ecluster} \\
 x_{\mathrm{cluster}} &=& \frac{\sum_{i=1}^{n} x_i e_i}{\sum_{i=1}^{n} e_i}, \label{eq:xcluster} \\
 y_{\mathrm{cluster}} &=& \frac{\sum_{i=1}^{n} y_i e_i}{\sum_{i=1}^{n} e_i}, ~\mathrm{and} \label{eq:ycluster} \\
  t_{\mathrm{cluster}} &=& \frac{ \sum_{i=1}^n t_i / \sigma_{t_i}^2 }{ \sum_{i=1}^n 1 / \sigma_{t_i}^2 }, \label{eq:tcluster}
\end{eqnarray}
where $n$ denotes the number of seeds in the cluster and $e_i$, $x_i$, $y_i$, and  $t_i$ are the energy, position, and time of the $i$ th seed, respectively.
The time is normalized by an energy-dependent resolution measured in test beam data as $\sigma_{t_i}= \sqrt{(5/e_i)^2 + (3.63/\sqrt{e_i})^2 + 0.13^2}$ ns, with $e_i$ in units of MeV~\cite{nim_iwai}.

 Clusters with more than one crystal and an energy deposit over 20~MeV were called ``photon clusters'' and used in the $\pi^0$ reconstruction. Single seed crystals that did not belong to any cluster were called ``single-crystal hits" and used as a signature of accidentals in the CsI calorimeter.
 
 The center of energy ($x_\mathrm{cluster}$, $y_\mathrm{cluster}$) was shifted from the actual incident position of the photon due to its nonzero incident angle on the surface of the CsI calorimeter, in conjunction with the depth of the main energy deposit. In addition, the energy of the cluster $E_\mathrm{cluster}$ deviated from the true photon energy due to the shower leakage outside the calorimeter. We corrected for these effects by using correction maps extracted from an MC simulation in advance. The resulting position resolution of the photon entrance position at the front face of the calorimeter was $3.7/\sqrt{E~\mathrm{[GeV]}}$~mm with a constant term of 2.1~mm.

\subsection{Event reconstruction}\label{sec:reconstruction}
In the analysis of $K_L \rightarrow 3\pi^0$, $K_L \rightarrow 2\pi^0$, and $K_L \rightarrow 2\gamma$ decays, the first step was to count the number of photon clusters. We required six or more photon clusters for $K_L \rightarrow 3\pi^0$, four or more photon clusters for $K_L \rightarrow 2\pi^0$, and two or more photon clusters for $K_L \rightarrow 2\gamma$ events. Next, we selected the six, four, or two photon clusters closest in time and started the process of reconstructing the $K_L$ vertex and its four-momentum.

\subsubsection{Vertex reconstruction} \label{sec:vertexreconstruction}
 For $K_L \rightarrow 2\gamma$ events, the event vertex was derived assuming that the two photon clusters have an invariant mass equal to the $K_L$ mass and originate from the beam axis at coordinates (0, 0, $Z_{\mathrm{vtx}}$).

The reconstruction of the event vertex for $K_L \rightarrow 3\pi^0$ and $K_L \rightarrow 2\pi^0$ events started by constraining pairs of photon clusters to have an invariant mass equal to the nominal $\pi^0$ mass. The calculation of the event vertex position ($X_{\mathrm{vtx}},~Y_{\mathrm{vtx}},~Z_{\mathrm{vtx}}$) and timing $T_{\mathrm{vtx}}$ follows the same principles as what was done in the E391a experiment~\cite{Ahn:2010fk}; it is outlined in the following. 

 \begin{figure}
  \centering
  \includegraphics[width=8cm, bb=0 0 1000 600, clip ]{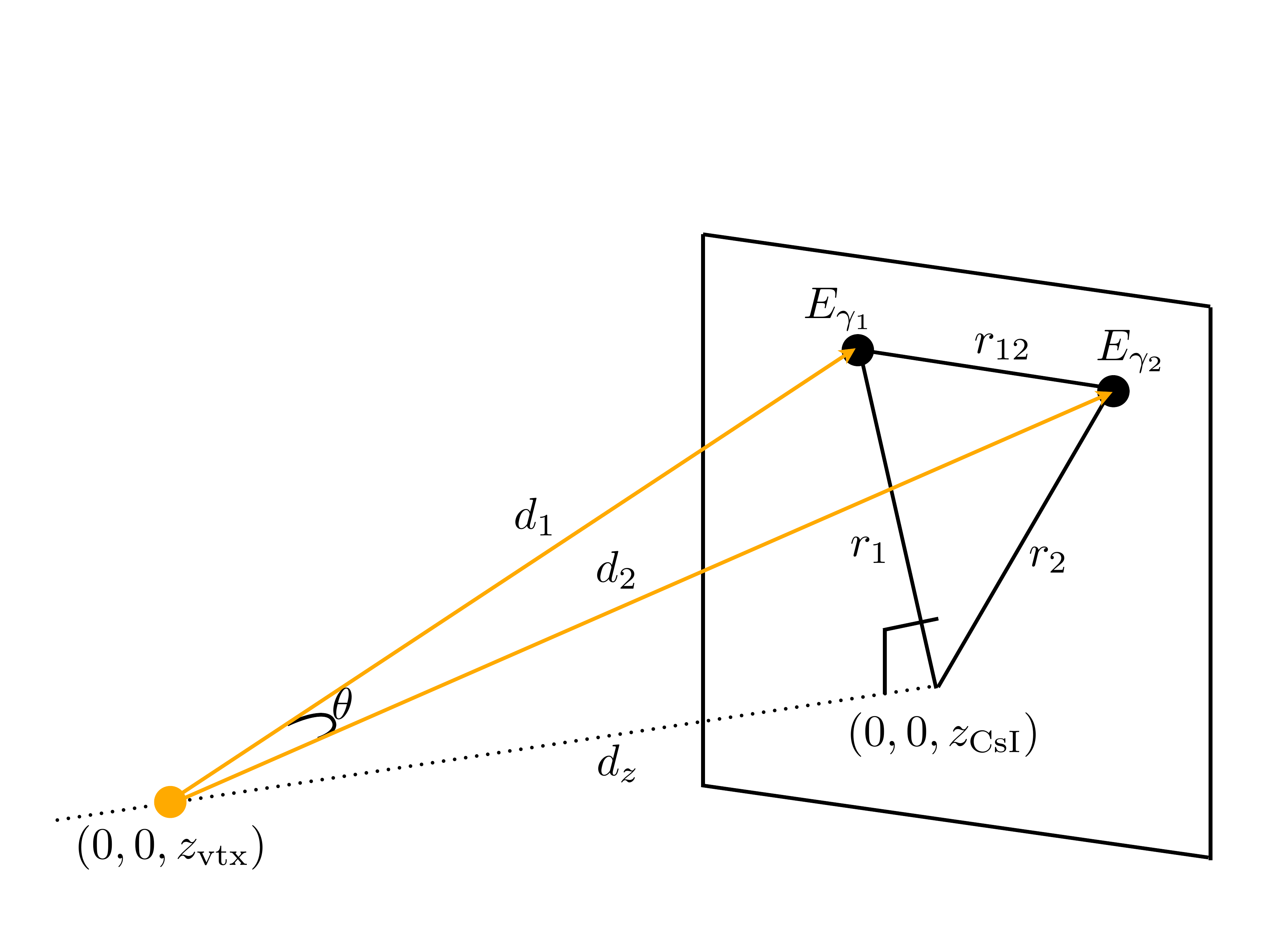}
  \caption{Graphical description of the $\pi^0$ reconstruction: the two yellow lines show the paths of the photons from the event vertex to their entrance position at the front face of the calorimeter and $\theta$ is their opening angle; $E_{\gamma_1}$, $E_{\gamma_2}$ are the energies deposited by the photons; $z_{\mathrm{CsI}}$ is the $z$ position of the front surface of the CsI calorimeter in detector coordinates; $d_z$ is the distance of the calorimeter from the event vertex; $d_1$, $d_2$ are the distances of the two photon hits from the event vertex; and $r_1$, $r_2$ are their projection at the front face of the calorimeter.}
  \label{fig:pi0rec}
 \end{figure}
 
 Figure \ref{fig:pi0rec} represents a graphical description of the $\pi^0$ reconstruction. Assuming that the invariant mass of the two photons is equal to the nominal $\pi^0$ mass $M_{\pi^0}$~\cite{pdg}, the following equation holds:
\begin{equation}
  {M_{\pi^0}}^2 = 2E_{\gamma{_1}} E_{\gamma{_2}} ( 1 - \cos \theta ),  \label{eq:costheta} 
\end{equation}
where $E_{\gamma{_1}}$ and $E_{\gamma{_2}}$ are the energies of the photons and $\theta$ is the angle between their directions. 
In addition, the following geometrical relations hold:
 \begin{eqnarray}
  {r_{12}}^2 &=& {d_1}^2 + {d_2}^2 - 2 d_1 d_2 \cos \theta,  \label{eq:r12} \\
  d_1 &=& \sqrt{{r_1}^2 + {d_z}^2}, \mathrm{~and} \\
  d_2 &=& \sqrt{{r_2}^2 + {d_z}^2}, \label{eq:d2}
 \end{eqnarray}
 where $r_{12}$ is the distance between the hit positions of the two photons at the front face of the calorimeter and $d_i$ ($r_i$) is the distance between the hit position of the $i$ th photon and the decay vertex ($z$-axis). From Eqs.~(\ref{eq:costheta})--(\ref{eq:d2}), one can calculate the distance of the $\pi^0$ from the front face of the calorimeter $d_z$, with an uncertainty derived from propagating the single photon energy and position resolutions. These in turn can be used to derive the position $Z$ and uncertainty $\sigma_{z}$ of the $\pi^0$ vertex in detector coordinates.

The six (four) photon clusters in the $K_L \rightarrow 3\pi^0$ ($K_L \rightarrow 2\pi^0$) samples can be paired to form three (two) $\pi^0$ in 15 (3) independent combinations. To find the correct photon pairing, we minimized the ``pairing variance'' $\chi_z^2$:
\begin{eqnarray}
 \chi_z^2 = \sum_{i=1}^{n} \frac{ (Z_i - \bar{Z})^2}{\sigma_{zi}^2}, \\
 \bar{Z} = \frac{\sum_{i=1}^{n} Z_i / \sigma_{zi}^2}{\sum_{i=1}^{n} 1 / \sigma_{zi}^2}, \label{eq:klzvertex}
\end{eqnarray}
where $n$ is the number of pairs in a given combination. We assigned the $\bar{Z}$ of the combination to the smallest $\chi_z^2$ as the $K_L$ decay position $Z_{\mathrm{vtx}}$.

The $K_L$ vertex position in the $x$-$y$ plane was calculated as the interpolated position between the T1 target and the center of energy on the CsI front face at the position $Z_{\mathrm{vtx}}$, assuming that the T1 target is a point source.

Once the three-dimensional vertex position was determined, the generated time of each photon at the vertex $T_j$ was calculated from the photon cluster time $t_j$ after correcting for the time-of-flight as $T_j = t_j - d_j/ c$, where $d_j$ is the distance between the event vertex and the calorimeter front face (see Fig.~\ref{fig:pi0rec}), and $c$ is the speed of light.  Finally, the time of the $K_L$ decay time $T_{\mathrm{vtx}}$ was defined as the weighted mean of the photon generated times:
\begin{equation}
 T_{\mathrm{vtx}} = \frac{\sum_{j=1}^{2n} T_j / \sigma_{t_j}^2}{\sum_{j=1}^{2n} 1 / \sigma_{t_j}^2}, \label{eq:vertextime}
\end{equation}
where $\sigma_{t_j}$ is the timing resolution of a cluster given by the sum in quadrature of the energy-dependent term, $3.8 / \sqrt{E_j~\mathrm{[MeV]}}$~ns, and the constant term 0.19~ns.

\subsubsection{Mass and momentum reconstruction} \label{sec:massreconstruction}
 So far we have assumed that the photon pairs in $K_L \rightarrow 3\pi^0$ and $K_L \rightarrow 2\pi^0$ decays have the nominal $\pi^0$ mass. With the reconstructed vertex position in hand, these constraints were removed and the four-momenta of the $\pi^0$ were recalculated by assuming the vertex position ($X_{\mathrm{vtx}}, ~Y_{\mathrm{vtx}}, ~Z_{\mathrm{vtx}}$). The four-momentum of the initial $K_L$ was obtained by summing over the four-momenta of the $\pi^0$. For the $K_L \rightarrow 2\gamma$ decay, the four-momentum of $K_L$ was calculated from the four-momenta of the two photons assuming the vertex position (0, 0, $Z_{\mathrm{vtx}}$).

\subsection{Event selection}\label{sec:selection}
  The selection of $K_L \rightarrow 3\pi^0$ events by requiring exactly six photon clusters in the CsI calorimeter resulted in a sample with 10\% background contamination in the MC simulation. A similar requirement of exactly four photon clusters for $K_L \rightarrow 2\pi^0$ events, and exactly two for $K_L \rightarrow 2\gamma$ events, resulted in signal-to-background ratios of 0.18\% and 0.25\%, respectively. Selection criteria (cuts) on the kinematics of the event and in the presence of extra particles were applied for each mode to improve the signal-to-background ratio, as described in Sects.~\ref{sec:kinematic} and \ref{sec:veto}. Most cuts are common to all modes, as summarized in Tables~\ref{tab:3pi0selection}, \ref{tab:2pi0selection}, and \ref{tab:2gammaselection}.

\subsubsection{Kinematic cuts}\label{sec:kinematic}
 \begin{itemize}
 \item{$\Delta T_{\mathrm{vtx}}$ cut} \mbox{} \\
 The difference between the time $T_j$ of each photon in Eq.~(\ref{eq:vertextime}) and the reconstructed vertex time $T_{\mathrm{vtx}}$ was required to be less than 3~ns. This cut reduced the contamination of accidentals in signal events.

 \item{($X_{\mathrm{min}},Y_{\mathrm{min}}$) cut} \mbox{} \\
 The position of the innermost photon was required to be outside an area of 120~mm $\times$ 120~mm from the CsI calorimeter center. This cut removed photons whose shower leaked into the beam hole.

\item{$R_{\mathrm{max}}$ cut} \mbox{} \\
 The position of the outermost photon was required to be inside a radius of 850~mm from the center of the CsI calorimeter. This cut removed photons whose shower leaked outside the calorimeter fiducial volume. In the $K_L \rightarrow 2\gamma$ analysis, the outermost photon was also required to have a minimum radius of 450~mm in order to remove background events from the $K_L \rightarrow 3\pi^0$ mode decaying near the front face of the CsI calorimeter.

\item{$E_{\mathrm{min}}$ cut} \mbox{} \\
 The energy of each photon was required to be larger than 50~MeV. This cut removed photon clusters with poor energy and position resolutions.

\item{$d_{\mathrm{min}}$ cut} \mbox{} \\
 The distance between the hit position of any two photons at the front face of the CsI calorimeter was required to be larger than 150~mm. This cut reduces the probability of misreconstructing a single photon into multiple clusters or multiple photons into a single cluster.

\item{$\chi^2_{\mathrm{shape}}$ cut} \mbox{} \\
In order to distinguish electromagnetic showers generated by single photons from showers generated by multiple photons or hadronic interactions, we defined the cluster shape variable:
\begin{equation}
 \chi^2_{\mathrm{shape}} = \frac{1}{N} \sum_{i=1}^N \left( \frac{E_i^{\mathrm{obs}} - E_i^{\mathrm{ref}}}{\sigma_i^{\mathrm{ref}}} \right)^2,
\end{equation}
where $N$ is the number of crystals involved in the cluster, $E_i^{\mathrm{obs}} (E_i^{\mathrm{ref}})$ denotes the observed (expected) energy deposit in each crystal, and $\sigma_i^{\mathrm{ref}}$ is the uncertainty on the expected energy deposit. The single-crystal energy deposit, and its uncertainty, were derived from an MC simulation by using photons of different incident energies and polar and azimuthal angles~\cite{dt_sato}. 
A cut of $ \chi^2_{\mathrm{shape}} < 5$ was used for the $K_L \rightarrow 2\pi^0$ and  $K_L \rightarrow 2\gamma$ events.

\item{$\Delta M_{\pi^0}$ cut} \mbox{} \\
 The reconstructed invariant mass of the $\pi^0$ was required to be within 10~$\mathrm{MeV/}c^2$ of the nominal $M_{\pi^0}$ for the $K_L\rightarrow3\pi^0$ mode and within 6~$\mathrm{MeV/}c^2$ for the $K_L\rightarrow2\pi^0$ mode. This cut rejected background events with mispaired photon clusters.

\item{$\Delta M_{K_L}$ cut} \mbox{} \\
 The reconstructed invariant mass for $K_L \rightarrow 2\pi^0$ events was required to be within 15~$\mathrm{MeV/}c^2$ of the nominal $K_L$ mass~\cite{pdg}. This cut was effective for removing $3\pi^0$ background events with two photons not reconstructed inside the CsI calorimeter fiducial volume.
   
\item{$P_T(K_L)$ cut} \mbox{} \\
 The reconstructed $K_L$ transverse momentum for the $K_L \rightarrow 2\gamma$ decay mode was required to be less than 50~MeV/$c$, higher than the 30-MeV/$c$ peak observed in the $P_T(K_L)$ distribution of simulated events. This cut rejected background events, in particular from the $K_L \rightarrow 3\pi^0$ mode, where some missing transverse momentum was carried away by undetected particles.

\item{$Z_{\mathrm{vtx}}$ cut} \mbox{} \\
 The reconstructed $K_L$ vertex was bounded to be between 2000~mm and 5400~mm in the detector coordinate for the $K_L \rightarrow 3\pi^0$ and the $K_L \rightarrow 2\pi^0$ modes, and between 2000~mm and 5000~mm for the $K_L \rightarrow 2\gamma$ mode.
 
\item{($X_{\mathrm{coe}},Y_{\mathrm{coe}}$) cut} \mbox{} \\
 The event center-of-energy, given by the energy-weighted $x$ and $y$ coordinates of all photons at the front face of the calorimeter, was required to be within a $60 \times 60~\mathrm{mm^2}$ area around the center of the calorimeter. This cut helped to reject background events with missing energy coming from undetected particles in the $K_L \rightarrow 2\pi^0$ mode.

\item{$\Sigma E_{1/2}$ cut} \mbox{} \\
 The sum of the energy deposited by photons in both the left and right halves of the CsI calorimeter was required to be above 350~MeV, above the trigger threshold of 307.5~MeV.
 
\end{itemize}

\subsubsection{Veto cuts}\label{sec:veto}
\begin{itemize}
 \item{CsI veto} \mbox{} \\
 In addition to measuring energies and positions of photons, the CsI calorimeter also served as a veto detector for extra photons. Events were removed if one or more photon clusters were present in addition to the six, four, and two photons for the $K_L \rightarrow 3\pi^0$, $K_L \rightarrow 2\pi^0$, and $K_L\rightarrow 2\gamma$ modes, respectively. The additional photon cluster was counted only if its timing was within $\pm10$~ns of the average time of the photon clusters used in the event reconstruction.

\item{CsI single-crystal hit veto} \mbox{} \\
  Events with ``single-crystal hits'' not included in any cluster were rejected. Fluctuations in the electromagnetic shower occasionally result in single crystal hits near a photon cluster. To avoid signal acceptance loss, the minimum energy threshold $E_{\mathrm{thr}}$ for a single-crystal hit was determined as a function of the distance from the closest photon cluster, $d$, as:
\begin{align}
 E_{\mathrm{thr}} &= (23.4 - 0.034~d\mathrm{[mm]})~\mathrm{MeV}  &\mathrm{for~} 100<d<600~\mathrm{mm}, \\
 E_{\mathrm{thr}} &= 3~\mathrm{MeV} &\mathrm{for~} d\ge 600~\mathrm{mm}.
\end{align}
Furthermore, the time of the single-crystal hit was required to be within $\pm10$~ns of the closest photon cluster. We did not veto events with any activity within 100~mm of the photon clusters.

\item{CV veto} \mbox{} \\
Penetrating charged particles deposit around 0.5~MeV in the CV. Based on this condition and the typical CV timing resolution (see Sect.~\ref{sec:detectors}), we defined a ``CV hit'' as any activity in a strip with a minimum energy deposit of 0.4~MeV in a time window of $\pm10$~ns from the event vertex $T_{\mathrm{vtx}}$, after correcting for particle time-of-flight. The energy and timing of the hit were calculated from the sum of the energy deposit and the average of the time measured at the two ends of the CV strip, respectively. By rejecting events with any hits in the CV, background events with charged particles were effectively suppressed.

\item{Inner MB veto} \mbox{} \\
The role of the inner MB veto was to veto the $K_L \rightarrow 3\pi^0$ background for the $K_L \rightarrow 2\pi^0$ and $K_L \rightarrow 2\gamma$ modes. The energy threshold was set at 5~MeV, and the time window, corrected for time-of-flight analogously to what was done for the CV, covered the interval [$-26,+34$]~ns. The inner MB timing resolution was 1.3~ns for a 5-MeV energy deposit. A wide time window was chosen to allow for time differences between the MB and the CsI calorimeter due to the MB length (5.5~m) along the beam direction. The energy, position, and time of a hit within the MB, after correcting for attenuation effects and the propagation delay in the WLSFs, were calculated using the signals from both the upstream and downstream ends of a MB module as:
\begin{eqnarray}
 t_{\mathrm{MB}} &=& \frac{t_{\mathrm{up}} + t_{\mathrm{down}} }{2}, \label{eq:mbtmod}\\
 z_{\mathrm{MB}} &=& \frac{ t_{\mathrm{up}} - t_{\mathrm{down}} }{2} \times v_{\mathrm{prop}}, ~\mathrm{and} \label{eq:mbzmod}\\
 E_{\mathrm{MB}} &=& \frac{e_{\mathrm{up}}}{ \exp \left[ - z_{\mathrm{MB}} / (\Lambda + \alpha z_{\mathrm{MB}}) \right] } + \frac{e_{\mathrm{down}}}{ \exp \left[ z_{\mathrm{MB}} / ( \Lambda - \alpha z_{\mathrm{MB}} ) \right] }, \label{eq:mbemod}
\end{eqnarray}
where $e_{\mathrm{up}} (e_{\mathrm{down}})$ is the visible energy and $t_{\mathrm{up}} (t_{\mathrm{down}})$ the hit time at the upstream (downstream) side, $z_{\mathrm{MB}}$ is the hit position along the beam axis with respect to the center of the MB module, $v_{\mathrm{prop}}=168.1$~mm/ns is the propagation velocity of light in the WLSF, and $\Lambda=4923$~mm and $\alpha=0.495$ parameterize the effects of the attenuation of the signal along the length of the MB.

\end{itemize}

\begin{table}
 \caption{Cuts used in the selection for the $K_L \rightarrow 3\pi^0$ mode} \label{tab:3pi0selection}
 \centering
 \begin{tabular}{lcc} \hline \hline
    Kinematic cut & Min. & Max. \\ \hline
   $\Delta T_{\mathrm{vtx}}$ & & 3~ns \\
  ($X_{\mathrm{min}},Y_{\mathrm{min}}$) & 120~mm\ & \\
    $R_{\mathrm{max}}$ & & 850~mm\\
    $E_{\mathrm{min}}$ & 50~MeV & \\
    $d_{\mathrm{min}}$ & 150~mm & \\
    $\Delta M_{\pi^0}$ & & 10~$\mathrm{MeV/}c^2$ \\
    $\Delta M_{K_L}$ & & 15~$\mathrm{MeV/}c^2$ \\
    $Z_{\mathrm{vtx}}$ & 2000~mm & 5400~mm \\ 
    $\Sigma E_{1/2}$ & 350~MeV & \\ \hline \hline
 Veto cut & Energy thresh. & Time window \\ \hline
    CsI & 20~MeV & $-$10~ns -- +10~ns \\ \hline \hline
 \end{tabular}
\end{table}

\begin{table}
 \caption{Cuts used in the selection for the $K_L \rightarrow 2\pi^0$ mode} \label{tab:2pi0selection}
 \centering
 \begin{tabular}{lcc} \hline \hline
    Kinematic cut & Min. & Max. \\ \hline
    $\Delta T_{\mathrm{vtx}}$ & & 3~ns \\
    ($X_{\mathrm{min}},Y_{\mathrm{min}}$) & 120~mm & \\
    $R_{\mathrm{max}}$  & & 850~mm \\
    $E_{\mathrm{min}}$ & 50~MeV & \\
    $d_{\mathrm{min}}$ & 150~mm & \\
    $\chi^2_{\mathrm{shape}}$ & & 5 \\
    $\Delta M_{\pi^0}$ & & 6~$\mathrm{MeV/}c^2$ \\
    $\Delta M_{K_L}$ & & 15~$\mathrm{MeV/}c^2$ \\
     $Z_{\mathrm{vtx}}$ & 2000~mm & 5400~mm \\
    $(X_{\mathrm{coe}},Y_{\mathrm{coe}})$  & & 60~mm\\ 
    $\Sigma E_{1/2}$ & 350~MeV & \\ \hline \hline
    Veto cut & Energy thresh. & Time window \\ \hline
    CsI & 20~MeV & $-$10~ns -- +10~ns \\
    CsI single-crystal hit & 3--20~MeV & $-$10~ns -- +10~ns \\
    CV & 0.4~MeV & $-$10~ns -- +10~ns \\
    Inner MB & 5~MeV & $-$26~ns -- +34~ns \\ \hline \hline
 \end{tabular}
\end{table}

\begin{table}
 \caption{Cuts used in the selection for the $K_L \rightarrow 2\gamma$ mode} \label{tab:2gammaselection}
 \centering
 \begin{tabular}{lcc} \hline \hline
    Kinematic cut & Min. & Max. \\ \hline
    $\Delta T_{\mathrm{vtx}}$ & & 3~ns \\
    ($X_{\mathrm{min}},Y_{\mathrm{min}}$) & 120~mm & \\
    $R_{\mathrm{max}}$  & 450~mm & 850~mm \\
    $\chi^2_{\mathrm{shape}}$ & & 5 \\  
    $P_T (K_L)$ & & 50~MeV/$c$ \\
    $Z_{\mathrm{vtx}}$ & 2000~mm & 5000~mm \\
    $\Sigma E_{1/2}$ & 350~MeV & \\ \hline \hline
    Veto cut & Energy thresh. & Time window \\ \hline
    CsI & 20~MeV& $-$10~ns -- +10~ns \\
    CsI single-crystal hit & 3--20~MeV & $-$10~ns -- +10~ns \\
    CV & 0.4~MeV & $-$10~ns -- +10~ns \\
    Inner MB & 5~MeV & $-$26~ns -- +34~ns \\ \hline \hline
 \end{tabular}
\end{table}

\section{Results}\label{sec:kl_flux}
 In this section we derive the result of the $K_L$ flux from the measured $K_L$ yield in each of the $K_L \rightarrow 3\pi^0$, $K_L \rightarrow 2\pi^0$, and $K_L \rightarrow 2\gamma$ samples, after normalizing to the known number of protons on target. The $K_L$ yield $Y$ is defined as $Y = N^{\mathrm{data}}_{\mathrm{rec}}/(A \times $BF) where $N^{\mathrm{data}}_{\mathrm{rec}}$ is the number of reconstructed data events after applying all the selection cuts and the background subtraction that is described in this section. $A$ is the product of the decay probability and the acceptance calculated as the ratio of MC events passing the selection cuts over the total number of $K_L$ that passed the beam exit and decayed to the given decay mode, and BF is the branching fraction for the specific decay mode listed in Table~\ref{tab:klbr}. In the following sections, the inputs to the yield measurement for each of the three modes are presented.

\subsection{$K_L$ yield for $K_L \rightarrow 3\pi^0$ mode}
\begin{figure}
  \centering
  \includegraphics[width=9cm, bb=0 0 570 360, clip]{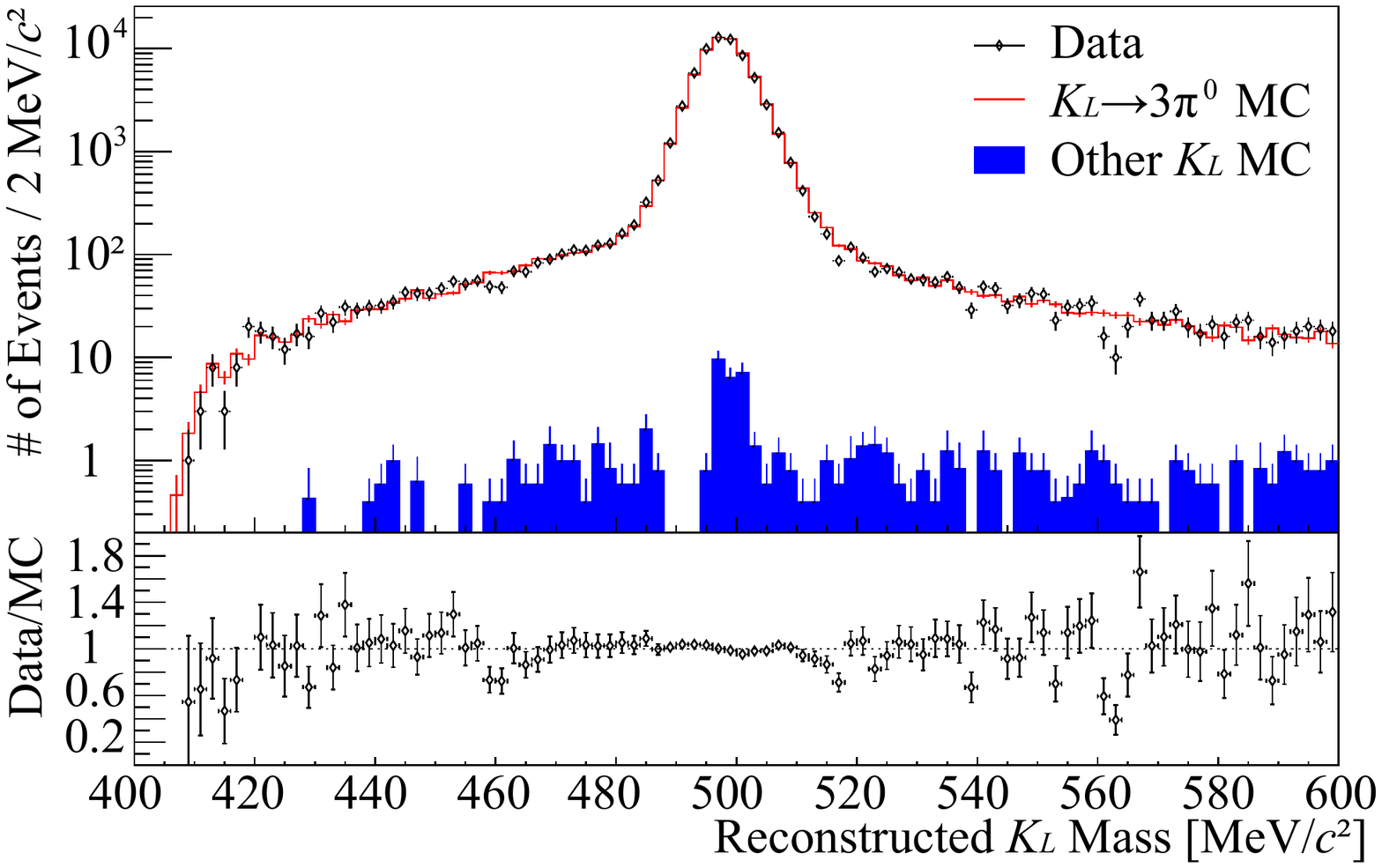}
  \caption{Reconstructed $K_L$ mass spectrum in the $K_L \rightarrow 3\pi^0$ data sample. The black points represent the experimental data, the open red histogram represents the $K_L \rightarrow 3\pi^0$ MC simulation, and the solid blue histogram represents the estimated contribution from the other $K_L$ decay modes. The lower plot shows the ratio between the experimental data and the MC simulation. The error bars reflect only the statistical uncertainties.}
  \label{fig:3pi0mass}
\end{figure}

 The reconstructed invariant mass distribution for $K_L \rightarrow 3\pi^0$ candidate events after applying all the selection cuts in Table~\ref{tab:3pi0selection}, except for the $\Delta M_{K_L}$ cut, is shown in Fig.~\ref{fig:3pi0mass}. The MC distributions in the figure are normalized so that the number of reconstructed events after applying all the selection cuts in the MC simulation with the background contamination is equal to that in the data. The background contamination was estimated to be less than 0.1\%, smaller than both the statistical uncertainty (0.39\%) and the uncertainty in the nominal BF (0.61\%~\cite{pdg}). The final number of reconstructed events in the data and the MC acceptance are listed in Table~\ref{tab:3pi0normalizaton}. The $K_L$ yield at the beam exit for the $K_L \rightarrow 3\pi^0$ mode was estimated to be $(7.835 \pm 0.033) \times 10^8$. The uncertainty includes only the statistical uncertainties in the data and MC inputs.

\begin{table}
 \caption{Number of reconstructed events in the data after applying all the selection cuts for the $K_L \rightarrow 3\pi^0$ mode and subtracting the background contamination. The MC acceptance for the signal and the dominant background contribution, before and after correcting for their BFs are also listed. The uncertainty in the acceptance reflects only the MC statistics. The last column shows the relative contribution of MC events for each decay mode.}
 \label{tab:3pi0normalizaton}
 \centering
  \begin{tabular}{lccc} \hline \hline
     $N_{\mathrm{rec}}^{\mathrm{data}}$ & $65,284 \pm 256$ & & \\ \hline \hline
     & $A$ & $A \times $BF & (\%) \\ \hline
     $K_L \rightarrow 3\pi^0$  & $(4.269 \pm 0.007) \times 10^{-4}$ &  $(8.332 \pm 0.013) \times 10^{-5}$ & 99.95 \\
    $K_L \rightarrow \pi^+ \pi^- \pi^0$ & $(1.74 \pm 0.19) \times 10^{-7}$ & $(2.18 \pm 0.23) \times 10^{-8}$ & 0.03 \\
    \hline \hline
 \end{tabular}
\end{table}

\subsection{$K_L$ yield for $K_L \rightarrow 2\pi^0$ mode}
\begin{figure}
  \centering
  \includegraphics[width=9cm, bb=0 0 580 360, clip]{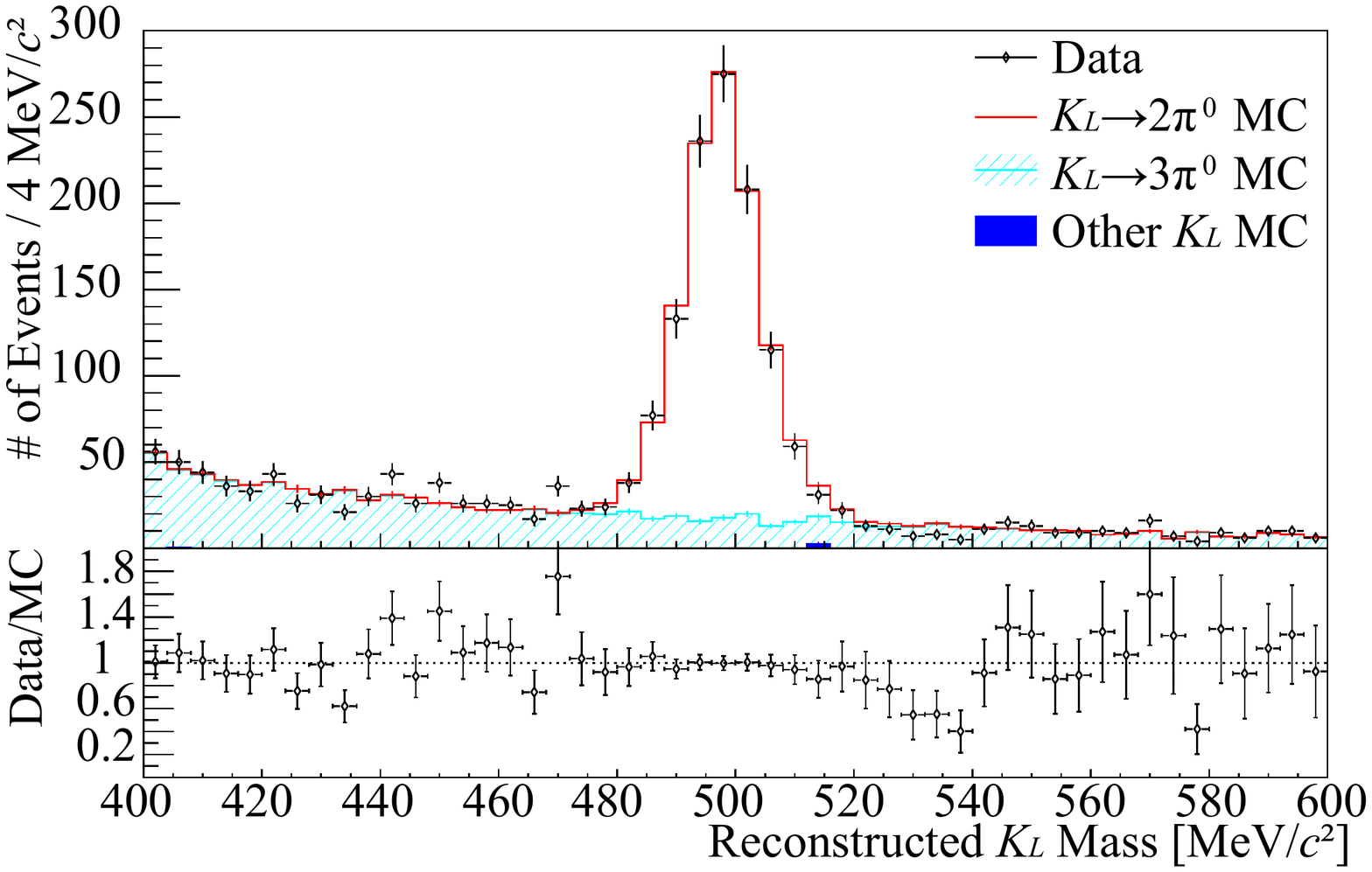}
  \caption{Reconstructed $K_L$ mass spectrum in the $K_L \rightarrow 2\pi^0$ selected data sample. The black points represent the experimental data, the open red histogram represents the $K_L \rightarrow 2\pi^0$ MC simulation, the light-blue histogram represents the estimated contribution from the $K_L \rightarrow 3\pi^0$ mode, and the solid blue histogram represents the contribution from the other $K_L$ decay modes. The lower plot shows the ratio between the experimental data and the MC simulation. The error bars reflect only the statistical uncertainties.}
  \label{fig:2pi0mass}
\end{figure}

The reconstructed invariant mass distribution for $K_L \rightarrow 2\pi^0$ candidate events after applying all the selection cuts in Table~\ref{tab:2pi0selection}, except for the $\Delta M_{K_L}$ cut, is shown in Fig.~\ref{fig:2pi0mass}. The MC distributions in the figure are normalized to the yield measured for the $K_L \rightarrow 3\pi^0$ mode. The background contamination was estimated by minimizing:
\begin{equation}
 \chi^2 = \sum_{i=1}^{N} \frac{(d_i - \alpha s_i - \beta b_i)^2}{ (\sigma_{d_i})^2 + (\alpha \sigma_{s_i})^2 + (\beta \sigma_{b_i})^2 } \label{eq:chisquare}
\end{equation}
with respect to the signal and background scale factors, $\alpha$ and $\beta$. Here, $N$ is the number of bins, $d_i \pm \sigma_{d_i}$ is the number of data events in the $i$ th bin and its statistical uncertainty, and $s_i \pm \sigma_{s_i}$ ($b_i \pm \sigma_{b_i}$) is the number of MC signal (background) events and their relative uncertainties normalized to the $K_L \rightarrow 3\pi^0$ flux result. The $\chi^2$ minimization returned $\alpha = 0.993 \pm 0.034$ and $\beta = 0.956 \pm 0.036$ with $\chi^2/\mathrm{ndf~(number~of~degrees~of~freedom)} = 53.3/48$. The calculated number of background events was $120.8 \pm 6.2$ out of 1129 total $K_L \rightarrow 2\pi^0$ candidate events.

The number of reconstructed signal events and the MC acceptance are summarized in Table~\ref{tab:2pi0normalizaton}. The $K_L$ yield at the beam exit using the $K_L \rightarrow 2\pi^0$ mode was estimated to be $(7.85 \pm 0.27) \times 10^8$. The uncertainty includes only the statistical uncertainties in the data and MC inputs. 

\begin{table}
 \caption{Number of reconstructed events in the data after applying all the selection cuts for the $K_L \rightarrow 2\pi^0$ mode and subtracting the background contamination. The MC acceptances for the signal and the dominant background contribution, before and after correcting for their respective BFs are also listed. The uncertainty in the acceptance reflects only the MC statistics. Only the backgrounds contributing more than 0.1\% are listed; their relative contributions are shown in the last column. }
  \label{tab:2pi0normalizaton}
 \centering
 \begin{tabular}{lccc} \hline \hline
     $N_{\mathrm{rec}}^{\mathrm{data}}$ & 1008.2 $\pm$ 34.2 & & \\ \hline \hline
     & $A$ & $A \times $BF &  (\%) \\ \hline
    $K_L \rightarrow 2\pi^0$  & $(1.487 \pm 0.012) \times 10^{-3}$ & $(1.284 \pm 0.011) \times 10^{-6}$ & 88.8 \\
    $K_L \rightarrow 3\pi^0$ & $(8.23 \pm 0.29) \times 10^{-7}$ &  $(1.606 \pm 0.056) \times 10^{-7}$ & 11.1 \\
    \hline \hline
 \end{tabular}
\end{table}

\subsection{$K_L$ yield for $K_L \rightarrow 2\gamma$ mode}
\begin{figure}
  \centering
  \includegraphics[width=9cm, bb=0 0 580 360, clip]{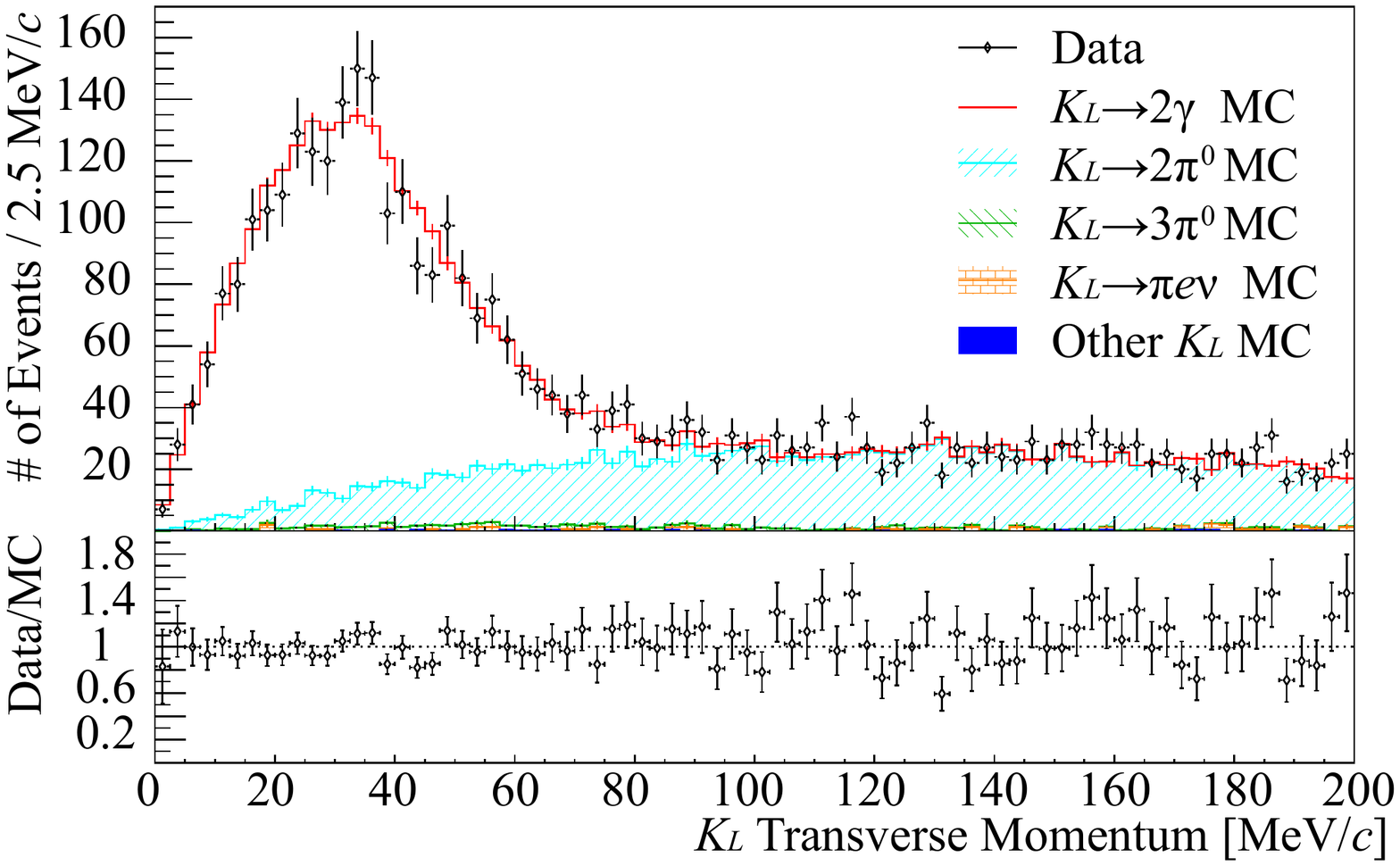}
  \caption{Reconstructed $K_L$ transverse-momentum spectrum in the $K_L \rightarrow 2\gamma$ selected data sample. The black points represent the experimental data, the open red histogram represents the $K_L \rightarrow 2\gamma$ MC simulation, and the light-blue, green, and orange histograms represent the estimated contribution from the $K_L\rightarrow 2\pi^0$, $K_L\rightarrow 3\pi^0$, and $K_{e3}$ decay modes, respectively. The blue histogram represents the other $K_L$ decay modes. The lower plot shows the ratio between the experimental data and the MC simulation. The error bars reflect only the statistical uncertainties.}
  \label{fig:2gammapt}
\end{figure}

 The reconstructed transverse-momentum distribution for $K_L \rightarrow 2\gamma$ candidate events after applying all the selection cuts in Table~\ref{tab:2gammaselection}, except for the $P_T(K_L)$ cut, is shown in Fig.~\ref{fig:2gammapt}. The MC distributions in the figure are normalized to the yield measured for the $K_L \rightarrow 3\pi^0$ mode.  Analogously to what was done in the $K_L \rightarrow 2\pi^0$ analysis, the background contamination in the candidate event sample was extracted by minimizing the $\chi^2$ in Eq.~(\ref{eq:chisquare}). The minimization returned $\alpha = 0.973 \pm 0.025$ and $\beta = 1.023 \pm 0.031$ with $\chi^2/\mathrm{ndf} = 67.1/78$. The calculated number of background events was $200.9 \pm 8.4$ out of 1890 total $K_L \rightarrow 2\gamma$ candidate events.
 
 The number of reconstructed signal events and the MC acceptances are summarized in Table~\ref{tab:2gammanormalizaton}. The $K_L$ yield at the beam exit using the $K_L \rightarrow 2\gamma$ mode was estimated as $(7.65 \pm 0.20) \times 10^8$. The uncertainty includes only the statistical uncertainties in the data and MC inputs. 
 
\begin{table}
 \caption{Number of reconstructed events in the data after applying all the selection cuts for the $K_L \rightarrow 2\gamma$ mode and subtracting the background contamination. The MC acceptances for the signal and the dominant background contributions, before and after correcting for their respective BFs are also listed. The uncertainty in the acceptance reflects only the MC statistics. Only backgrounds contributing more than 0.1\% are listed; their relative contributions are shown in the last column. }
  \label{tab:2gammanormalizaton}
 \centering
 \begin{tabular}{lccc} \hline \hline
    $N_{\mathrm{rec}}^{\mathrm{data}}$ & 1689.1 $\pm$ 44.3 & & \\ \hline \hline
    & $A$ & $A \times $BF & (\%) \\ \hline
    $K_L \rightarrow 2\gamma$ & $(4.035 \pm 0.020) \times 10^{-3}$ & $(2.207 \pm 0.011) \times 10^{-6}$ & 89.8 \\
    $K_L \rightarrow 3\pi^0$ & $(1.133 \pm 0.034) \times 10^{-6}$ &  $(2.212 \pm 0.066) \times 10^{-7}$ & 9.0 \\
    $K_L \rightarrow 2\pi^0$ & $(2.58 \pm 0.16) \times 10^{-5}$ & $(2.23 \pm 0.14) \times 10^{-8}$ & 0.9 \\
    $K_L \rightarrow \pi^{\pm} e^{\mp} \nu$ & $(1.4 \pm 0.4) \times 10^{-8}$ & $(5.68 \pm 2.15) \times 10^{-9}$ & 0.2 \\
    \hline \hline
 \end{tabular}
\end{table}

\subsection{$K_L$ flux}\label{sec:klfluxfinal}
 The measured $K_L$ yield for each decay mode is summarized in Table~\ref{tab:statonly}. The three results are consistent within the statistical uncertainties. Their weighted average gives the final $K_L$ flux of $Y_{\mathrm{ave}}=(7.831 \pm 0.032) \times 10^8$. The uncertainty includes only the statistical uncertainty.
 
 \begin{table}
 \caption{$K_L$ yield obtained for the $K_L \rightarrow 3\pi^0$, $K_L \rightarrow 2\pi^0$, $K_L \rightarrow 2\gamma$ decays, and their weighted average. The last column gives the relative yield for each decay mode with respect  to the average.}
  \label{tab:statonly}
 \centering
 \begin{tabular}{lcc} \hline \hline
    Mode & $K_L$ yield ($\times 10^8$) & Relative yield \\ \hline
    $K_L \rightarrow 3\pi^0$ & $7.835 \pm 0.033$ & 1.001 $\pm$ 0.004 \\
    $K_L \rightarrow 2\pi^0$ & $7.85 \pm 0.27$ & 1.002 $\pm$ 0.034 \\
    $K_L \rightarrow 2\gamma$ & $7.65 \pm 0.20$ & 0.977 $\pm$ 0.026 \\  \hline
    Average & $7.831 \pm 0.032$ & \\ \hline \hline
 \end{tabular}
\end{table}

The absolute $K_L$ yield in units of $2 \times 10^{14}$~POT, which is the J-PARC design value for the number of protons on target per spill, was calculated from the $K_L$ yield as: 
\begin{eqnarray}
 \frac{\mathrm{Number~of}~K_L}{2 \times 10^{14}~ \mathrm{POT}} &=& \frac{Y_{\mathrm{ave}}}{\varepsilon_{\mathrm{trg}}} \times  \frac{(2 \times 10^{14} )}{ N_{\mathrm{POT}}} \notag \\
  &=& (4.188\pm 0.017) \times 10^7, \label{eq:statresult}
\end{eqnarray}
where 
$\varepsilon_{\mathrm{trg}} = 11.15$\% is the KOTO trigger efficiency corrected for losses due to the DAQ dead time, and $N_{\mathrm{POT}}$ is the total number of protons on target collected during the run.  The uncertainty includes only the statistical uncertainty of the three modes.

\subsection{Systematic uncertainties}\label{sec:systematics}
 In this subsection, we describe the sources and estimation methods for the systematic uncertainties that affect the flux measurement.
 
\begin{figure}
  \centering
  \includegraphics[width=15cm, bb=20 0 520 270, clip]{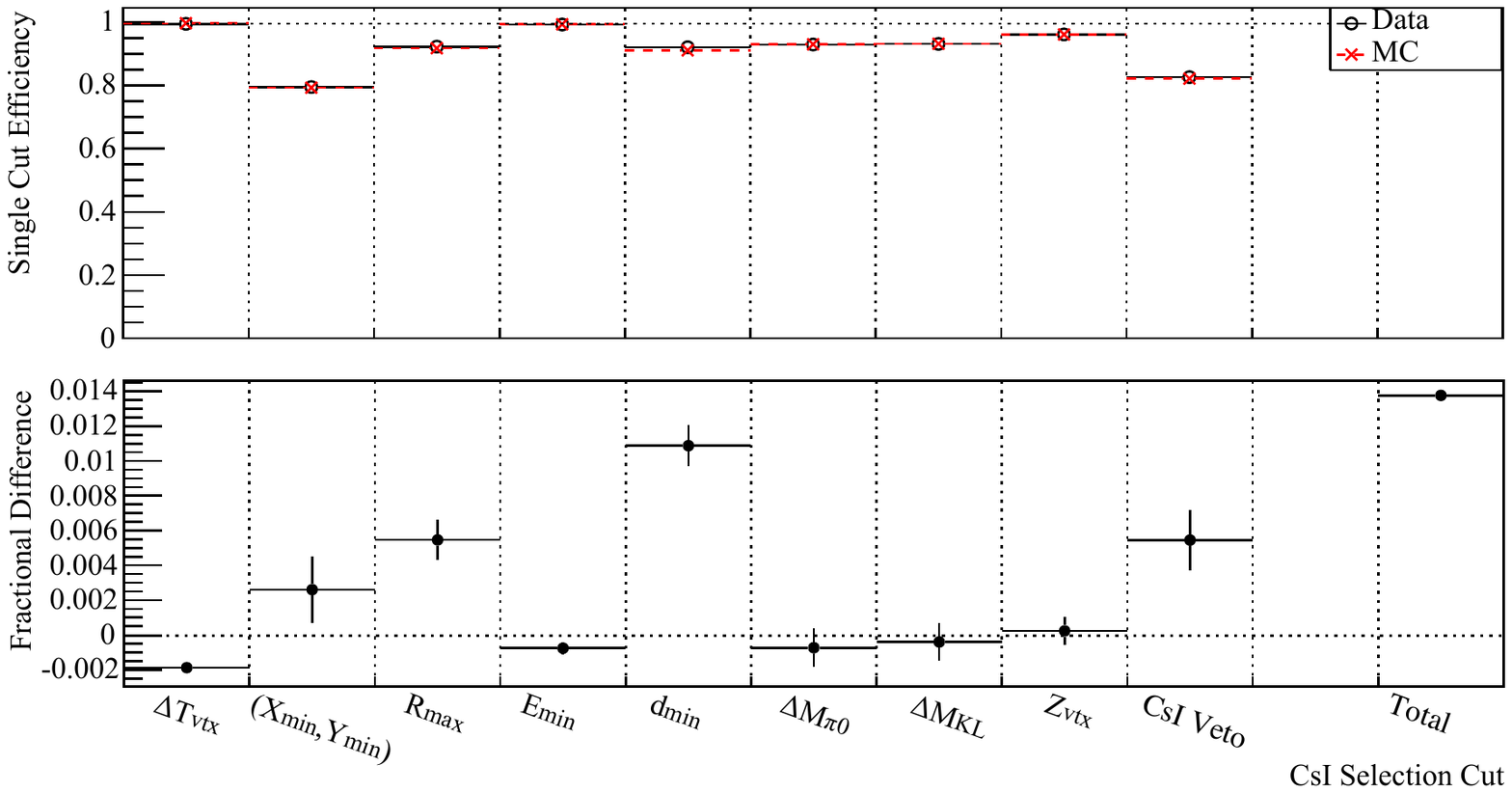}
  \caption{Single-cut efficiency (top) and the fractional difference (bottom) for each CsI selection cut, including the extra cluster veto cut, used in the $K_L \rightarrow 3\pi^0$ analysis.}
  \label{fig:3pi0cuteffectiveness}
\end{figure}

\begin{figure}
  \centering
  \includegraphics[width=15cm, bb=20 0 520 280, clip]{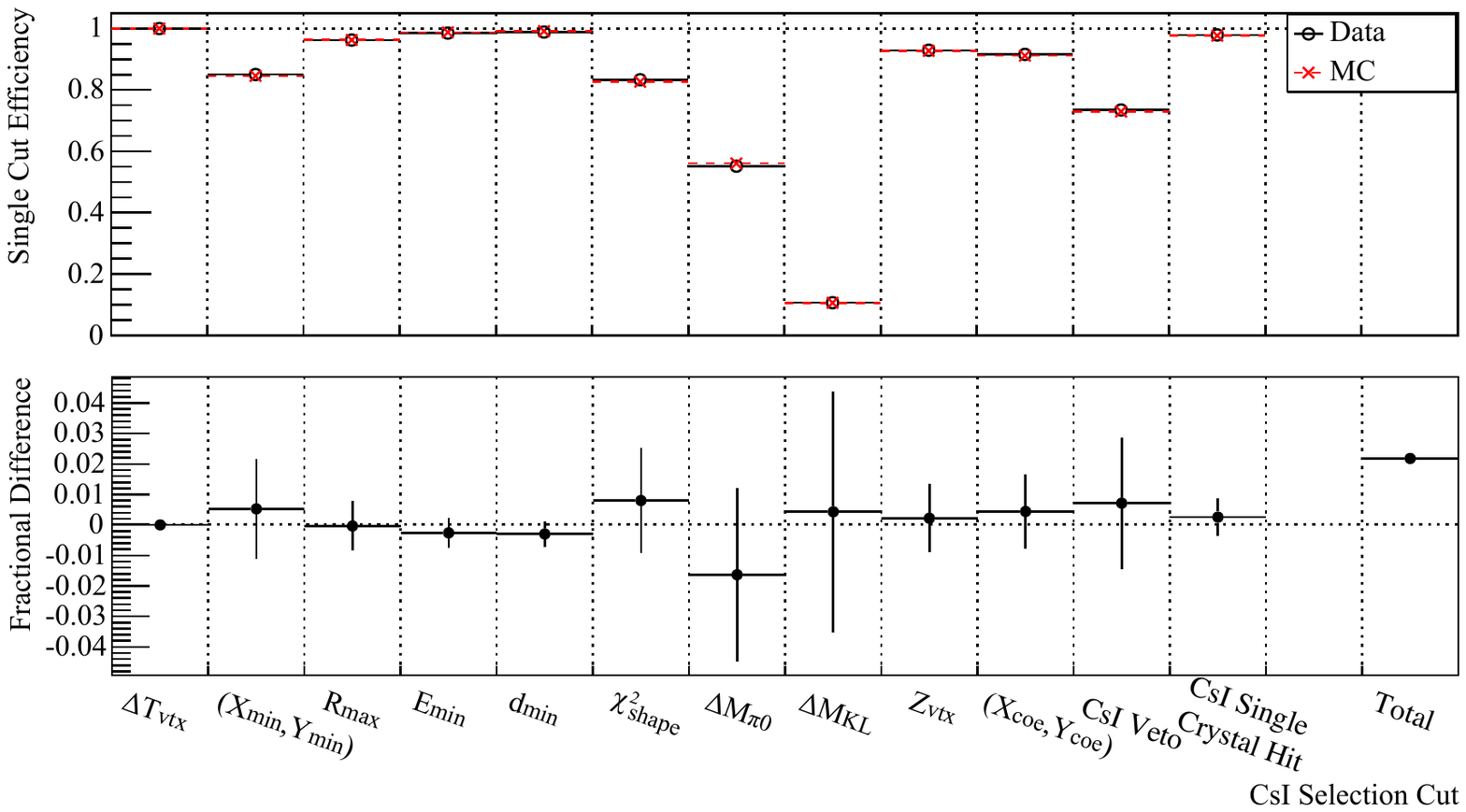}
  \caption{Single-cut efficiency (top) and the fractional difference (bottom) for each CsI selection cut, including the extra cluster veto cut, used in the $K_L \rightarrow 2\pi^0$ analysis. }
  \label{fig:2pi0cuteffectiveness}
\end{figure}

\begin{figure}
  \centering
  \includegraphics[width=15cm, bb=20 0 520 270, clip]{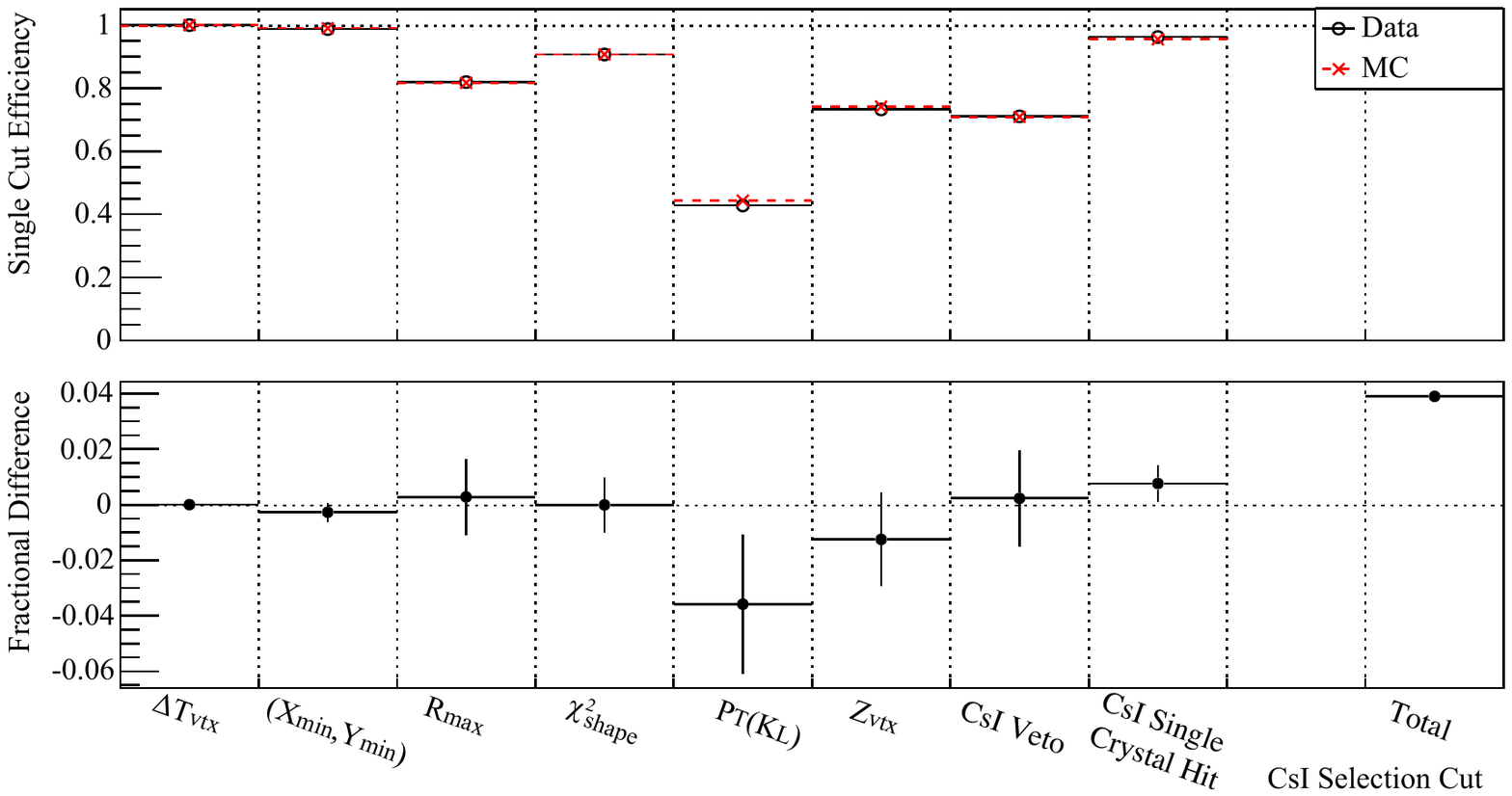}
  \caption{Single-cut efficiency (top) and the fractional difference (bottom) for each CsI selection cut, including the extra cluster veto cut, used in the $K_L \rightarrow 2\gamma$ analysis.}
  \label{fig:2gammacuteffectiveness}
\end{figure}

The uncertainty in the signal acceptance coming from the modeling of the calorimeter was estimated by comparing the effectiveness of each kinematic and CsI veto cut in data and MC. We calculated the single-cut fractional difference, defined as the ratio of data and MC efficiency for a given cut after all others have been applied. Figure~\ref{fig:3pi0cuteffectiveness} shows the single-cut efficiency of data and MC and their fractional differences for all the CsI-based cuts in the $K_L \rightarrow 3\pi^0$ analysis, including the extra cluster veto cut. By summing in quadrature all the fractional differences, we obtained a total systematic uncertainty coming from the modeling of the CsI calorimeter of 1.38\%. The systematic uncertainties for the other two modes were calculated in an analogous way and found to be 2.18\% and 3.90\% for the $K_L \rightarrow 2\pi^0$ analysis and the $K_L \rightarrow 2\gamma$ analysis, respectively. Figures~\ref{fig:2pi0cuteffectiveness} and \ref{fig:2gammacuteffectiveness} show the single-cut efficiencies and their fractional differences for $K_L \rightarrow 2\pi^0$ and $K_L \rightarrow 2\gamma$, respectively.
  	
The systematic uncertainty in the acceptance from the modeling of the veto detectors' response has two components: uncertainty from accidental losses and from backsplash losses. The accidental loss was due to accidental activity depositing energy in any veto detector in coincidence with a real photon in the CsI calorimeter. The backsplash loss was caused by particles belonging to a photon electromagnetic shower escaping the front of the CsI calorimeter and generating secondary activity in the veto detectors. The uncertainty in the modeling of both sources was studied by comparing the efficiency of the veto detector cuts in data and MC for $K_L \rightarrow 3\pi^0$ events, for which no CV or inner MB veto cuts were applied. The change in $K_L$ yield was measured to be 1.00\% after applying the inner MB veto cut and 0.65\% after applying the CV cut. These differences were added as systematic uncertainties in the $K_L \rightarrow 2\pi^0$ and $K_L \rightarrow 2\gamma$ selections.

Other sources of systematic uncertainties were estimated by changing the offline energy threshold from 350~MeV to the online trigger threshold value of 307.5~MeV. The resulting change in the number of selected events was taken as the systematic uncertainty of the $\Sigma E_{1/2}$  cut. The $K_L$ yield calculation used the Particle Data Group (PDG) branching fraction central values for the three neutral decay modes; the uncertainties on the central values reported in Table~\ref{tab:klbr} have been considered as a source of systematic uncertainty. Finally, the conversion factor from counts in the SEC to proton intensity used in Sect.~\ref{sec:runcondition} originates the 0.39\% uncertainty in the $N_{\mathrm{POT}}$ which is common to all the decay modes.

\begin{figure}
  \centering
  \includegraphics[width=10cm, bb=0 0 550 370, clip]{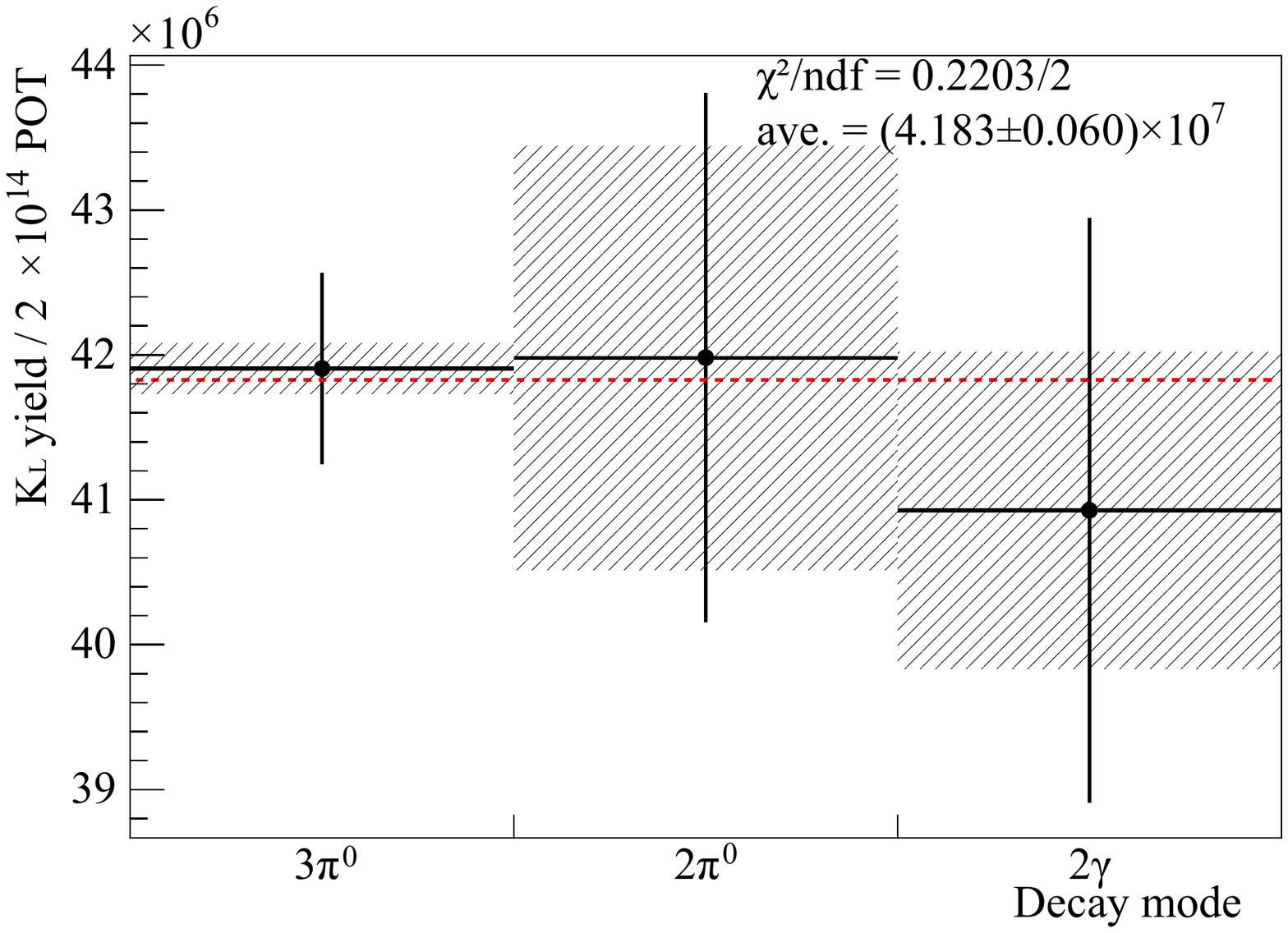}
  \caption{$K_L$ flux result for each of the $K_L \rightarrow 3\pi^0$, $K_L \rightarrow 2\pi^0$, and $K_L \rightarrow 2\gamma$ modes. The hatched area shows the statistical uncertainty and the vertical bar shows the quadratic sum of the statistical and mode-dependent systematic uncertainties. The horizontal dashed line shows the weighted average value.}
  \label{fig:result}
\end{figure}

 All the sources of systematic uncertainties are summarized in Table~\ref{tab:systotal}. The largest source is the modeling of the CsI calorimeter. All the other uncertainties are smaller than the statistical uncertainty for a given mode. The final uncertainty has been calculated by adding in quadrature all the statistical and mode-dependent systematic uncertainties for a single mode, taking their weighted average, and adding in quadrature the mode-independent $N_{\mathrm{POT}}$ uncertainty. From Eq.~(\ref{eq:statresult}), the final $K_L$ flux result was $(4.183 \pm 0.017_{\mathrm{stat.}} \pm 0.059_{\mathrm{sys.}}) \times 10^7$ $K_L$ per $2 \times 10^{14}$ protons on target. Figure~\ref{fig:result} compares the $K_L$ flux result separately for the three modes.

\begin{table}
 \caption{Summary of the systematic uncertainties. The first five sources are added in quadrature to obtain the mode-dependent systematic uncertainty of each decay channel. The $N_{\mathrm{POT}}$ uncertainty is independent of the three modes.} \label{tab:systotal}
 \centering
 \begin{tabular}{lccc} \hline \hline
 	Source  &  $K_L \rightarrow 3\pi^0$ & $K_L \rightarrow 2\pi^0$ & $K_L \rightarrow 2\gamma$ \\ \hline
	 CsI calorimeter modeling & 1.38\% & 2.18\% & 3.90\% \\
	 Main Barrel modeling & -- & 1.00\% & 1.00\% \\
	 Charged Veto modeling & -- & 0.65\% & 0.65\% \\
	 $\Sigma E_{1/2}$ cut  & 0.24\% & 0.41\% & 0.04\% \\
	 PDG branching fraction~\cite{pdg} & 0.61\% & 0.69\% & 0.73\% \\ \hline
	 Mode-dependent & 1.53\% & 2.61\% & 4.14\% \\ \\
	 Mode-independent ($N_{\mathrm{POT}}$) & \multicolumn{3}{c}{0.39\%} \\
	\hline \hline

 \end{tabular}
\end{table}
 
\section{Discussion}\label{sec:discussion}
The result of this paper can be compared to the previous $K_L$ flux measurement of $(4.19 \pm 0.09^{+0.47}_{-0.44}) \times 10^7$ per $2 \times 10^{14}$ POT, obtained from data taken in a dedicated beam survey run in February 2010~\cite{Shiomi2012264}. The experimental running conditions during the two measurements are summarized in Table~\ref{tab:twomeasurements}. Although the T1 targets were made of different materials (Au in 2013 vs. Pt in 2010), they have the same proton interaction length $\lambda_p$ within 3\%~\cite{anp}.

 \begin{table}
 \caption{Comparison of target material and target thickness, both in units of mm and proton interaction length $\lambda_p$, for the two KOTO $K_L$ flux measurements. The last two columns summarize the typical beam power conditions and the $K_L$ decay modes used for each measurement.} \label{tab:twomeasurements}
 \centering
 \begin{tabular}{lccccc} \\ \hline
	\multirow{2}{*}{Period} & \multicolumn{3}{c}{Target} & \multirow{2}{*}{MR beam power} & \multirow{2}{*}{Measured mode} \\ 
	 & material & thickness & $\lambda_p$ & & \\ \hline
	Feb. 2010 & Pt & 60.0~mm & 0.658 & 1~kW, 1.5~kW & $K_L \rightarrow \pi^+ \pi^- \pi^0$ \\ \\
	 & & & & &  $K_L \rightarrow 3\pi^0$ \\
	Jan. 2013 & Au & 66.0~mm & 0.640 & 15~kW & $K_L \rightarrow 2\pi^0$ \\
	 & & (1-mm slits included) & & &  $K_L \rightarrow 2\gamma$ \\ \hline
 \end{tabular}
\end{table}

Scaling the $K_L$ flux measured here to the J-PARC design values of 300~kW for the beam power and 0.7-s spill duration every 3.3~s \cite{Yamanaka01012012}, we predict a $K_L$ flux 100 times larger than that available in the previous $K_L \rightarrow \pi^0 \nu \overline{\nu}$ search experiment. Together with the upgrades to the detector, the experiment should reach the sensitivity of the standard model prediction for the $K_L \rightarrow \pi^0 \nu \overline{\nu}$ search over a period of three Snowmass years ($3 \times 10^7$~s).

\section{Conclusion}\label{sec:conclusion}
In this paper, we have described the $K_L$ flux measurement for the KOTO experiment with data taken during a detector commissioning run in January 2013. The measurement was done by using three $K_L$ neutral decay modes: $K_L \rightarrow 3\pi^0$, $K_L \rightarrow 2\pi^0$, and $K_L \rightarrow 2\gamma$. The results for the three decay modes agreed with each other within the statistical uncertainties. Systematic uncertainties were estimated based on the reproducibility of the data selection efficiency in the Monte Carlo simulation. The final $K_L$ flux was $(4.183 \pm 0.017 \pm 0.059) \times 10^7$ per $2 \times 10^{14}$ protons on target, where the first uncertainty was statistical and the second was systematic. This result is in agreement with a previous measurement done by the KOTO Collaboration during a dedicated beam survey run in February 2010.

\section*{Acknowledgments}

We would like to express our gratitude to all members of the J-PARC accelerator and Hadron Beam groups for their support and for providing stable beam operations. We also thank the KEK Central Computer System for providing the computing power which allowed us to handle the huge amount of data. This research was supported by the High Energy Accelerator Research Organization (KEK), the Ministry of Education, Culture, Sports, Science, and Technology (MEXT), the Japan Society for the Promotion of Science (JSPS) KAKENHI Grant Numbers 18071006, 10J00474, and 23224007, the United States Department of Energy, National Science Council/Ministry of Science and Technology in Taiwan, and the National Research Foundation of Korea (2012R-1A2A2A004554 and  2013K1A3A7A06056592(Center of Korean J-PARC Users)). The first author was supported by a Grant-in-Aid for JSPS Fellows.

% can use a bibliography generated by BibTeX as a .bbl file
% BibTeX documentation can be easily obtained at:
% http://www.ctan.org/tex-archive/biblio/bibtex/contrib/doc/
%\bibliographystyle{ptephy}
%\bibliographystyle{plainnat}

%\bibliography{KLYield}
%
% once the .bbl file has been generated then place the text in your article.
%\providecommand{\noopsort}[1]{}\providecommand{\singleletter}[1]{#1}%

\end{document}